\documentclass[11pt,letterpaper]{article}
\usepackage[T1]{fontenc}
\usepackage{amsmath}
\usepackage{amsfonts}
\usepackage{amssymb}
\usepackage{amsthm}
\usepackage{graphicx}
\graphicspath{{img/}}
\usepackage{booktabs}
\usepackage{hyperref}
\usepackage{breakcites}
\usepackage{multicol}
\usepackage{bbm}
\usepackage[ruled,vlined]{algorithm2e}
\usepackage{adjustbox}
\usepackage{multirow}
\usepackage{subfig}
\usepackage{pgfplots}
\usepackage{makecell}

\usetikzlibrary{decorations.pathreplacing}
\usetikzlibrary{3d,decorations.text,shapes.arrows,positioning,fit,backgrounds, positioning}
\usetikzlibrary{matrix}

\tikzset{%
	every neuron/.style={
		circle,
		draw,
		minimum size=0.5cm
	},
	neuron missing/.style={
		draw=none, 
		scale=2,
		text height=.333cm,
		execute at begin node=\color{black}$\vdots$
	},
}

\tikzset{pics/fake box/.style args={
		#1 with dimensions #2 and #3 and #4}{
		code={
			\draw[black,ultra thin,fill=#1]  (0,0,0) coordinate(-front-bottom-left) to
			++ (0,#3,0) coordinate(-front-top-right) --++
			(#2,0,0) coordinate(-front-top-right) --++ (0,-#3,0) 
			coordinate(-front-bottom-right) -- cycle;
			\draw[black,ultra thin,fill=#1] (0,#3,0)  --++ 
			(0,0,#4) coordinate(-back-top-left) --++ (#2,0,0) 
			coordinate(-back-top-right) --++ (0,0,-#4)  -- cycle;
			\draw[black,ultra thin,fill=#1!80!black] (#2,0,0) --++ (0,0,#4) coordinate(-back-bottom-right)
			--++ (0,#3,0) --++ (0,0,-#4) -- cycle;
			\path[black,decorate,decoration={text effects along path,text={CONV}}] (#2/2,{2+(#3-2)/2},0) -- (#2/2,0,0);
		}
}}

\tikzset{pics/fake box2/.style args={
		#1 with dimensions #2 and #3 and #4}{
		code={
			\draw[black,ultra thin,fill=#1]  (0,0,0) coordinate(-front-bottom-left) to
			++ (0,#3,0) coordinate(-front-top-right) --++
			(#2,0,0) coordinate(-front-top-right) --++ (0,-#3,0) 
			coordinate(-front-bottom-right) -- cycle;
			\draw[black,ultra thin,fill=#1] (0,#3,0)  --++ 
			(0,0,#4) coordinate(-back-top-left) --++ (#2,0,0) 
			coordinate(-back-top-right) --++ (0,0,-#4)  -- cycle;
			\draw[black,ultra thin,fill=#1!80!black] (#2,0,0) --++ (0,0,#4) coordinate(-back-bottom-right)
			--++ (0,#3,0) --++ (0,0,-#4) -- cycle;
			\path[black,decorate,decoration={text effects along path,text={LATENT}}] (#2/2,{2.75+(#3-2.75)/2},0) -- (#2/2,0,0);
		}
}}

\tikzset{pics/fake boxx/.style args={
		#1 with dimensions #2 and #3 and #4}{
		code={
			\draw[black,ultra thin,fill=#1]  (0,0,0) coordinate(-front-bottom-left) to
			++ (0,#3,0) coordinate(-front-top-right) --++
			(#2,0,0) coordinate(-front-top-right) --++ (0,-#3,0) 
			coordinate(-front-bottom-right) -- cycle;
			\draw[black,ultra thin,fill=#1] (0,#3,0)  --++ 
			(0,0,#4) coordinate(-back-top-left) --++ (#2,0,0) 
			coordinate(-back-top-right) --++ (0,0,-#4)  -- cycle;
			\draw[black,ultra thin,fill=#1!80!black] (#2,0,0) --++ (0,0,#4) coordinate(-back-bottom-right)
			--++ (0,#3,0) --++ (0,0,-#4) -- cycle;
			\path[black,decorate,decoration={text effects along path,text={INPUT}}] (#2/2,{2+(#3-2)/2},0) -- (#2/2,0,0);
		}
}}

\tikzset{pics/fake boxo/.style args={
		#1 with dimensions #2 and #3 and #4}{
		code={
			\draw[black,ultra thin,fill=#1]  (0,0,0) coordinate(-front-bottom-left) to
			++ (0,#3,0) coordinate(-front-top-right) --++
			(#2,0,0) coordinate(-front-top-right) --++ (0,-#3,0) 
			coordinate(-front-bottom-right) -- cycle;
			\draw[black,ultra thin,fill=#1] (0,#3,0)  --++ 
			(0,0,#4) coordinate(-back-top-left) --++ (#2,0,0) 
			coordinate(-back-top-right) --++ (0,0,-#4)  -- cycle;
			\draw[black,ultra thin,fill=#1!80!black] (#2,0,0) --++ (0,0,#4) coordinate(-back-bottom-right)
			--++ (0,#3,0) --++ (0,0,-#4) -- cycle;
			\path[black,decorate,decoration={text effects along path,text={OUTPUT}}] (#2/2,{2+(#3-2)/2},0) -- (#2/2,0,0);
		}
}}

\tikzset{pics/fake embedding/.style args={
		#1 with dimensions #2 and #3 and #4}{
		code={
			\draw[black,ultra thin,fill=#1]  (0,0,0) coordinate(-front-bottom-left) to
			++ (0,#3,0) coordinate(-front-top-right) --++
			(#2,0,0) coordinate(-front-top-right) --++ (0,-#3,0) 
			coordinate(-front-bottom-right) -- cycle;
			\draw[black,ultra thin,fill=#1] (0,#3,0)  --++ 
			(0,0,#4) coordinate(-back-top-left) --++ (#2,0,0) 
			coordinate(-back-top-right) --++ (0,0,-#4)  -- cycle;
			\draw[black,ultra thin,fill=#1!80!black] (#2,0,0) --++ (0,0,#4) coordinate(-back-bottom-right)
			--++ (0,#3,0) --++ (0,0,-#4) -- cycle;
			\path[black,decorate,decoration={text effects along path,text={EMBEDDING}}] (#2/2,{4+(#3-4)/2},0) -- (#2/2,0,0);
		}
}}

\tikzset{pics/fake embeddingo/.style args={
		#1 with dimensions #2 and #3 and #4}{
		code={
			\draw[black,ultra thin,fill=#1]  (0,0,0) coordinate(-front-bottom-left) to
			++ (0,#3,0) coordinate(-front-top-right) --++
			(#2,0,0) coordinate(-front-top-right) --++ (0,-#3,0) 
			coordinate(-front-bottom-right) -- cycle;
			\draw[black,ultra thin,fill=#1] (0,#3,0)  --++ 
			(0,0,#4) coordinate(-back-top-left) --++ (#2,0,0) 
			coordinate(-back-top-right) --++ (0,0,-#4)  -- cycle;
			\draw[black,ultra thin,fill=#1!80!black] (#2,0,0) --++ (0,0,#4) coordinate(-back-bottom-right)
			--++ (0,#3,0) --++ (0,0,-#4) -- cycle;
			\path[black,decorate,decoration={text effects along path,text={OUTPUT}}] (#2/2,{3+(#3-3)/2},0) -- (#2/2,0,0);
		}
}}

\tikzset{pics/fake fc/.style args={
		#1 with dimensions #2 and #3 and #4}{
		code={
			\draw[black,ultra thin,fill=#1]  (0,0,0) coordinate(-front-bottom-left) to
			++ (0,#3,0) coordinate(-front-top-right) --++
			(#2,0,0) coordinate(-front-top-right) --++ (0,-#3,0) 
			coordinate(-front-bottom-right) -- cycle;
			\draw[black,ultra thin,fill=#1] (0,#3,0)  --++ 
			(0,0,#4) coordinate(-back-top-left) --++ (#2,0,0) 
			coordinate(-back-top-right) --++ (0,0,-#4)  -- cycle;
			\draw[black,ultra thin,fill=#1!80!black] (#2,0,0) --++ (0,0,#4) coordinate(-back-bottom-right)
			--++ (0,#3,0) --++ (0,0,-#4) -- cycle;
			\path[black,decorate,decoration={text effects along path,text={FC}}] (#2/2,{1+(#3-1)/2},0) -- (#2/2,0,0);
		}
}}

\tikzset{pics/fake unroll/.style args={
		#1 with dimensions #2 and #3 and #4}{
		code={
			\draw[black,ultra thin,fill=#1]  (0,0,0) coordinate(-front-bottom-left) to
			++ (0,#3,0) coordinate(-front-top-right) --++
			(#2,0,0) coordinate(-front-top-right) --++ (0,-#3,0) 
			coordinate(-front-bottom-right) -- cycle;
			\draw[black,ultra thin,fill=#1] (0,#3,0)  --++ 
			(0,0,#4) coordinate(-back-top-left) --++ (#2,0,0) 
			coordinate(-back-top-right) --++ (0,0,-#4)  -- cycle;
			\draw[black,ultra thin,fill=#1!80!black] (#2,0,0) --++ (0,0,#4) coordinate(-back-bottom-right)
			--++ (0,#3,0) --++ (0,0,-#4) -- cycle;
			\path[black,decorate,decoration={text effects along path,text={UNROLL}}] (#2/2,{1+(#3-1)/2},0) -- (#2/2,0,0);
		}
}}

\tikzset{pics/fake roll/.style args={
		#1 with dimensions #2 and #3 and #4}{
		code={
			\draw[black,ultra thin,fill=#1]  (0,0,0) coordinate(-front-bottom-left) to
			++ (0,#3,0) coordinate(-front-top-right) --++
			(#2,0,0) coordinate(-front-top-right) --++ (0,-#3,0) 
			coordinate(-front-bottom-right) -- cycle;
			\draw[black,ultra thin,fill=#1] (0,#3,0)  --++ 
			(0,0,#4) coordinate(-back-top-left) --++ (#2,0,0) 
			coordinate(-back-top-right) --++ (0,0,-#4)  -- cycle;
			\draw[black,ultra thin,fill=#1!80!black] (#2,0,0) --++ (0,0,#4) coordinate(-back-bottom-right)
			--++ (0,#3,0) --++ (0,0,-#4) -- cycle;
			\path[black,decorate,decoration={text effects along path,text={ROLL}}] (#2/2,{2+(#3-2)/2},0) -- (#2/2,0,0);
		}
}}

\tikzset{pics/fake roll/.style args={
		#1 with dimensions #2 and #3 and #4}{
		code={
			\draw[black,ultra thin,fill=#1]  (0,0,0) coordinate(-front-bottom-left) to
			++ (0,#3,0) coordinate(-front-top-right) --++
			(#2,0,0) coordinate(-front-top-right) --++ (0,-#3,0) 
			coordinate(-front-bottom-right) -- cycle;
			\draw[black,ultra thin,fill=#1] (0,#3,0)  --++ 
			(0,0,#4) coordinate(-back-top-left) --++ (#2,0,0) 
			coordinate(-back-top-right) --++ (0,0,-#4)  -- cycle;
			\draw[black,ultra thin,fill=#1!80!black] (#2,0,0) --++ (0,0,#4) coordinate(-back-bottom-right)
			--++ (0,#3,0) --++ (0,0,-#4) -- cycle;
			\path[black,decorate,decoration={text effects along path,text={ROLL}}] (#2/2,{2+(#3-2)/2},0) -- (#2/2,0,0);
		}
}}

\tikzset{pics/fake boxhorizontal/.style args={
		#1 with dimensions #2 and #3 and #4}{
		code={
			\draw[black,ultra thin,fill=#1]  (0,0,0) coordinate(-front-bottom-left) to
			++ (0,#3,0) coordinate(-front-top-right) --++
			(#2,0,0) coordinate(-front-top-right) --++ (0,-#3,0) 
			coordinate(-front-bottom-right) -- cycle;
			\draw[black,ultra thin,fill=#1] (0,#3,0)  --++ 
			(0,0,#4) coordinate(-back-top-left) --++ (#2,0,0) 
			coordinate(-back-top-right) --++ (0,0,-#4)  -- cycle;
			\draw[black,ultra thin,fill=#1!80!black] (#2,0,0) --++ (0,0,#4) coordinate(-back-bottom-right)
			--++ (0,#3,0) --++ (0,0,-#4) -- cycle;
			\path[black,decorate,decoration={text effects along path,text={EMBEDDING}}] (0.25,#3/2,0) -- ({#2},#3/2,0);
		}
}}

\tikzset{pics/fake empty/.style args={
		#1 with dimensions #2 and #3 and #4}{
		code={
			\draw[black,ultra thin,fill=#1]  (0,0,0) coordinate(-front-bottom-left) to
			++ (0,#3,0) coordinate(-front-top-right) --++
			(#2,0,0) coordinate(-front-top-right) --++ (0,-#3,0) 
			coordinate(-front-bottom-right) -- cycle;
			\draw[black,ultra thin,fill=#1] (0,#3,0)  --++ 
			(0,0,#4) coordinate(-back-top-left) --++ (#2,0,0) 
			coordinate(-back-top-right) --++ (0,0,-#4)  -- cycle;
			\draw[black,ultra thin,fill=#1!80!black] (#2,0,0) --++ (0,0,#4) coordinate(-back-bottom-right)
			--++ (0,#3,0) --++ (0,0,-#4) -- cycle;
		}
}}

\usepackage{import}
\usetikzlibrary{quotes,arrows.meta}
\usetikzlibrary{positioning}

\def\edgecolor{rgb:blue,4;red,1;green,4;black,3}

\usepackage{Ball}
\usepackage{Box}
\usepackage{RightBandedBox}

\usetikzlibrary{positioning}
\usetikzlibrary{3d} 

\pgfplotsset{compat=1.17}
\usetikzlibrary{calc}

\tikzstyle{arrow} = [thick,->,>=stealth]
\usetikzlibrary{trees,shapes,decorations, arrows, fit, calc}
\usetikzlibrary{patterns.meta}
\makeatletter
\DeclareRobustCommand{\rvdots}{%
	\vbox{
		\baselineskip4\p@\lineskiplimit\z@
		\kern-\p@
		\hbox{.}\hbox{.}\hbox{.}
}}
\makeatother
\usetikzlibrary{matrix,calc}

\usepackage[margin=1.00in]{geometry}

\title{A representation-learning approach for insurance pricing with images}

\author{Christopher Blier-Wong\thanks{Corresponding author, \href{mailto:cblierwo@uwaterloo.ca}{cblierwo@uwaterloo.ca}},
Luc Lamontagne and Etienne Marceau\\ Université Laval, Québec, Canada}

\date{May 13, 2023}  

\begin{document}

\maketitle

\begin{abstract}
Unstructured data are a promising new source of information that insurance companies may use to understand their risk portfolio better and improve the customer experience. However, these novel data sources are difficult to incorporate into existing ratemaking frameworks due to the size and format of the unstructured data. In this paper, we propose a framework to use street view imagery within a generalized linear model. To do so, we use representation learning to extract an embedding vector containing useful information from the image. This embedding is dense and low-dimensional, making it appropriate to use within existing ratemaking models. We find that there is useful information included in street view imagery to predict the frequency of claims for certain types of perils. This model can be used as-is in a ratemaking framework but also opens the door to future empirical research on attempting to extract the causal effect from images that lead to increased or decreased predicted claim frequencies. Throughout, we discuss the practical difficulties (technical and social) of using this type of data for insurance pricing. 
\end{abstract}


\section{Introduction}

\subsection{Motivation}

Images, as a novel data source, can intervene at many places within property and casualty insurance applications. For instance, they may improve the quoting process by filling in some fields in the quoting questionnaire (number of stories, material type of the facade, presence of garage), either by an insurance agent or automatically with an artificial intelligence system. Going a step further, one may also use images directly within a ratemaking model to investigate whether they can be predictive of future claim counts or claim severity. In a claims management process, images may accelerate the handling procedure: if a customer provides an image of the damaged house or vehicle, one may estimate (manually by a claims adjuster or by an artificial intelligence system) the cost of repair or replacement of the damaged goods; this will lead to closing claims at a faster rate. All of the above examples, if done correctly, enhance/simplify the customer experience, improve actuarial fairness or reduce operating expenses.

For an insurance company, ratemaking plays a crucial role. Traditional ratemaking models rely on structured data such as policyholder demographics, coverage details and loss history. However, the recent surge in the availability of unstructured data, particularly images, has opened up new possibilities for enhancing ratemaking models and improving risk assessment. One research question we seek to answer in this paper is \textit{Are there useful information within street view imagery to predict the frequency or the severity of home insurance claims?} Our main methodological contribution is to propose a framework based on representation learning that will enable us to answer this research question. We use state-of-the-art image recognition techniques to extract meaningful information from images, which we then use to predict the frequency and severity of insurance claims. This offers a significant advancement in ratemaking, allowing for more accurate and nuanced risk assessments, ultimately resulting in fairer pricing for policyholders.

Concrete examples of uses of images for risk management include \cite{biffis2017satellite}, where the authors use satellite data to model weather risk using precipitation variability indices. More recently, in an actuarial context, \cite{zhu2023deep} combines economic and weather data to predict a production index for crop yield. From a claims management perspective, the authors of  \cite{doshi2023vehicle} predict the cost of repairing a car after an accident when provided an image of a damaged vehicle. Street view imagery (SVI) has proven useful for many applications; see \cite{biljecki2021street} for a review in urban analytics and geographic information science. As an example, \cite{fang2021synthesizing} use SVI to produce land-use classification and land-use mapping, critical tools for urban planning and environmental research. In \cite{chen2022deep} and \cite{blanc2022caracterisation}, the authors use SVI to perform information extraction from images to automate the data-collection process, improving the quoting process for customers. To the best of our knowledge, the current work is the first to use images to predict claim frequency or claim severity. Such a study is now feasible due to recent advances in image models, their performance for practical situations and the interest of insurance companies to stay ahead of the competition by adopting data-driven strategies in ratemaking and risk selection. 

Our goal in this paper is to propose a framework that uses images as inputs to ratemaking models. One problem we will face is that insurance data has a low signal-to-noise ratio \cite{wuthrich2023isotonic}, and insurance datasets typically do not have millions of examples as in image models. For this reason, we will not be able to use state-of-the-art image models directly within our framework but use a method based on representation learning. A secondary goal of this paper is to perform an empirical study to determine if there is useful information within SVI to predict home insurance losses. An answer in the affirmative will encourage researchers and practitioners in actuarial science to improve our framework and understand better what causal elements from SVI contribute to increased or decreased prediction of risk. 

\subsection{Related works}\label{ss:related}

This work fits within the machine learning literature in actuarial science; see \cite{richman2020ai}, \cite{richman2020aia} or \cite{blierwong2021machine} for early reviews. In particular, we use representation learning to determine if there is useful information in images for insurance pricing. For an overview of representation learning from an actuarial perspective, we refer the reader to \cite{blierwong2021rethinking}. In particular, \cite{richman2020ai} have advocated for using entity embeddings to transform categorical data into dense vectors; this idea was also suggested in \cite{blierwong2021rethinking, shi2022nonlifea, embrechts2022recent, wuthrich2023statistical}. In \cite{delong2023use}, the authors suggest initializing the parameters associated with the entity embeddings by first training an autoencoder. In \cite{avanzi2023machine}, the authors propose a model called GLMMNet to include mixed effects within the embedding layer. 

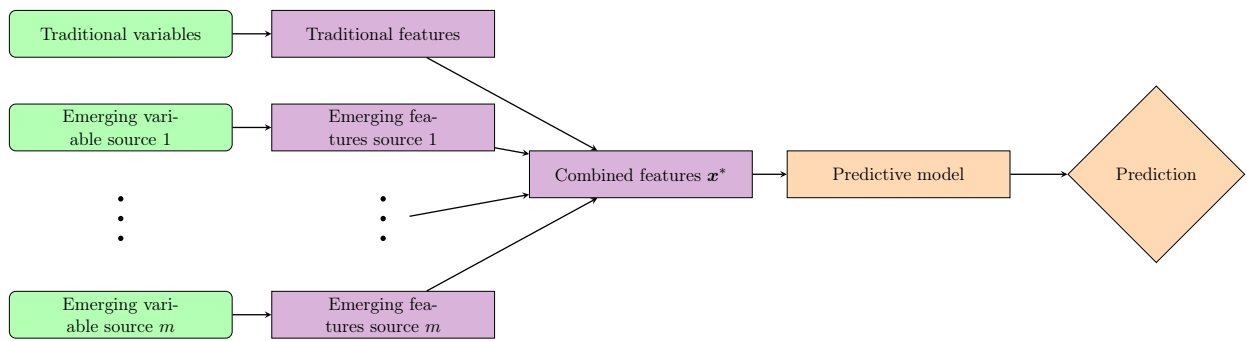
\begin{figure}[ht]
	\centering
	\resizebox{\textwidth}{!}{
		\begin{tikzpicture}[node distance=2cm]
			\node (data) [rectangle, rounded corners, minimum width=4.5cm, minimum height=1cm,text centered, text width = 4.5cm, draw=black, fill=green!30] {Traditional variables};
			\node (emerging1) [rectangle, rounded corners, minimum width=4.5cm, minimum height=1cm,text centered, text width = 4.5cm, draw=black, fill=green!30,below of=data] {Emerging variable source 1};
			\node (dots1) [below of = emerging1, scale = 3] {\rvdots};
			\node (emerging2) [rectangle, rounded corners, minimum width=4.5cm, minimum height=1cm,text centered, text width = 4.5cm, draw=black, fill=green!30,below of=dots1] {Emerging variable source $m$} ;
			
			\node (representation0) [rectangle, minimum width=4.5cm, minimum height=1cm, text centered, draw=black, text width = 4.5cm, fill=violet!30, right of = data, xshift=3.6cm] {Traditional features};
			\node (representation1) [rectangle, minimum width=4.5cm, minimum height=1cm, text centered, draw=black, text width = 4.5cm, fill=violet!30,right of = emerging1, xshift=3.6cm] {Emerging features source 1};
			\node (dots2) [right of = dots1, scale = 3, xshift=1.2cm] {\rvdots};
			\node (representation2) [rectangle, minimum width=4.5cm, minimum height=1cm, text centered, draw=black, text width = 4.5cm, fill=violet!30,right of = emerging2, xshift=3.6cm] {Emerging features source $m$};
			
			\node (representationc) [rectangle, minimum width=4.5cm, minimum height=1cm, text centered, draw=black, text width = 4.5cm, fill=violet!30,right of = representation1, xshift = 3.5cm, yshift=-1cm] {Combined features $\boldsymbol{x}^*$};
			
			\node (predict) [rectangle, minimum width=4.5cm, minimum height=1cm, text centered, draw=black, text width = 4.5cm, fill=orange!30, right of = representationc, xshift= 3.5cm] {Predictive model};
			\node (response) [diamond, minimum width=3cm, minimum height=3cm, text centered, text width=3cm, draw=black, fill=orange!30, right of = predict, xshift = 3.5cm] {Prediction};
			
			\draw [arrow] (data) -- (representation0);
			\draw [arrow] (emerging1) -- (representation1);
			\draw [arrow] (emerging2) -- (representation2);
			\draw [arrow] (representation0) -- (representationc);
			\draw [arrow] (representation1) -- (representationc);
			\draw [arrow] (dots2) -- (representationc);
			\draw [arrow] (representation2) -- (representationc);
			\draw [arrow] (representationc) -- (predict);
			\draw [arrow] (predict) -- (response);

	\end{tikzpicture}}
	\caption{Predictive learning in the unsupervised representation learning framework.}\label{fig:unsupervised}
\end{figure}

What sets our approach apart from the remaining literature is that we train our embeddings in an unsupervised way, meaning that we do not use the insurance loss data to construct the representations. We summarise the predictive learning framework proposed in \cite{blierwong2021rethinking} in Figure \ref{fig:unsupervised}. This framework lets one combine traditional actuarial variables with emerging variables such as spatial, image and textual data, within a simple predictive model (such as a GLM) for insurance pricing. Applications of the unsupervised representation learning framework in actuarial science include \cite{blierwong2022geographic}, which uses census data organized in a spatial way to construct geographic embeddings of postal codes to predict homeowners insurance claim frequency and \cite{xu2022bertbased}, who use pre-trained BERT models to construct representations of textual descriptions of claim data to predict the expected mean payment for truck warranty. The advantage of unsupervised representation learning is that we may construct complicated models for training the representations using large datasets of high-dimensional information like textual, image and spatial data and that these models do not overfit the task of interest, which in our case refers to predicting the future claim frequency/severity for an insurance contract. 

\subsection{Outline of paper}

The remainder of this paper is structured as follows. We present the general framework for insurance ratemaking using unsupervised embeddings in Section \ref{sec:unsupervised-framework}. In Section \ref{sec:data}, we present the SVI data that we will use to construct the representations in this paper, including the steps we take to prepare and clean the data. In Section \ref{sec:construct}, we construct representations of images starting from pre-trained image models. We start with a construction method that does not involve much effort and gradually increases the flexibility to determine when the model is useful enough for practical uses. In Section \ref{sec:application}, we use the embeddings constructed in the previous section to construct frequency and severity models using a real insurance dataset, and we conclude in Section \ref{sec:conclusion}.

\section{Unsupervised representation learning framework}\label{sec:unsupervised-framework}

A key step of our framework is to project SVI data into a low-dimensional vectorial format that captures useful features for insurance applications. In this section, we explain how one may use external data relevant to insurance to \textit{guide} a model to capture the useful parts of an image and let go of useless information. The general approach is inspired by word embeddings in natural language processing: instead of using one-hot encodings of words, it may be more convenient to use word representations that capture syntactic and semantic word relationships \cite{turian2010word, mikolov2013distributed, bengio2014representation}. The general framework for unsupervised representation-learning in an actuarial context is described in detail in \cite{blierwong2021rethinking}. In particular, it contains a general discussion of the data types, the intuition behind the representation learning framework and examples of applications in actuarial science. That framework has been applied to spatial data in \cite{blierwong2022geographic}. Also, the authors of \cite{lee2020actuarial} and \cite{xu2022bertbased} use a framework that can be considered as a special case of the one described in \cite{blierwong2021rethinking} with textual data. 

As explained in \cite{blierwong2021rethinking}, we believe that the unsupervised representation learning approach is well suited for insurance pricing due to the nature of insurance data (limited number of observations and low signal-to-noise ratio). However, from our experience, it only performs well if the resulting representation is related to the task of interest: adapting the embeddings to the insurance domain ensures that the insights generated by the models are meaningful and actionable for risk analysis. For this reason, we train our unsupervised embeddings on a task similar to our regression task of interest. Selecting appropriate related tasks is a vital step in constructing useful embeddings. Incorporating relevant information in the modelling process results in more accurate risk assessments and pricing, ultimately leading to fairer premiums for policyholders and better financial stability for insurers. 

An intuitive interpretation of the framework is that we aim to construct proxies of insurance-related information from SVI such that these proxies will be useful for ratemaking. These proxies are intermediate representations in a large image model trained on insurance-related tasks. The closer the insurance-related tasks are to predicting claim frequency or severity, the more useful the proxies should be. The representations should satisfy every principle related to rating variables; in particular, they should not rely on any protected attribute, such as the race or gender of people appearing in the SVI. 

Our approach is in the view of few-shot learning or unsupervised multitask learning in natural language processing  \cite{radford2019language, brown2020language} or image classification \cite{tian2020rethinking}. In few-shot learning, one attempts to design machine learning models that can effectively learn valuable information from a small amount of data and make accurate predictions or classifications. Few-shot learners sometimes start with task-agnostic or general multi-task methods, which are not tailored to any specific task. They are designed to capture broad patterns and information across various data types. Following this initial phase, they may undergo a process of fine-tuning, where they are specifically adjusted and optimized for more relevant and targeted tasks. This two-step process ensures that the model is easily adaptable to new tasks and can easily be guided to perform well on new specific tasks of interest. In an image classification context, the authors of \cite{tian2020rethinking} suggest that finding a good starting representation, followed by a linear classifier, outperforms other state-of-the-art few-shot learning models.

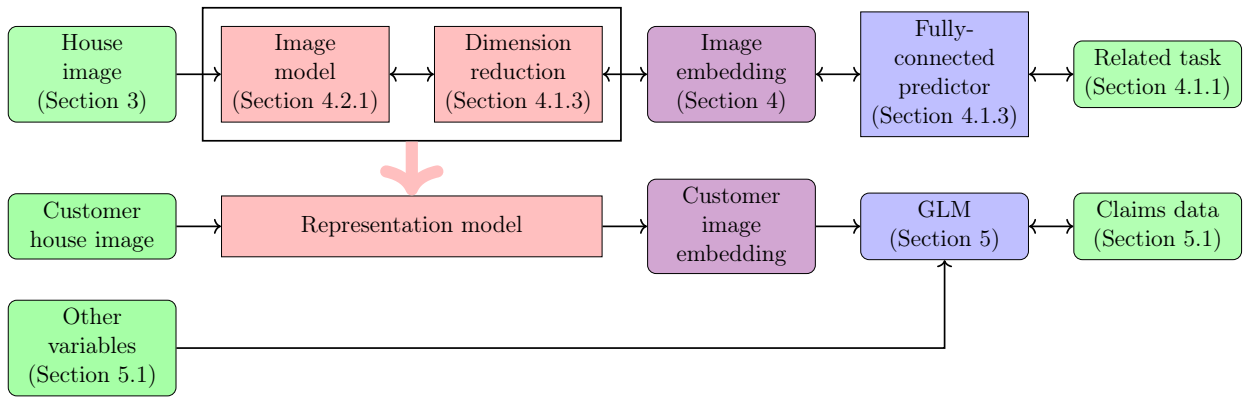
\begin{figure}[ht]
	\centering
	\resizebox{\textwidth}{!}{
		\begin{tikzpicture}[node distance=3.5cm]
			\node (data) [rectangle, rounded corners, minimum width=2.5cm, minimum height=1cm,text centered, text width = 2.5cm, draw=black, fill=green!30] {House \\ image \\ (Section \ref{sec:data})};	
			
			\node (encoder1) [rectangle, minimum width= 2.5cm, minimum height=1cm,text centered, text width = 2.5cm, draw=black, fill=red!25, right of = data] {Image \\ model  \\ (Section \ref{ss:backbone-image-model})};
			
			\node (encoder2) [rectangle, minimum width= 2.5cm, minimum height=1cm,text centered, text width = 2.5cm, draw=black, fill=red!25, right of = encoder1] {Dimension reduction\\ (Section \ref{ss:complete-approach})};
			
			\node (embedding) [rectangle, rounded corners, minimum width=2.5cm, minimum height=1cm,text centered, text width = 2.5cm, draw=black, fill=violet!35, right of = encoder2] {Image embedding \\ (Section \ref{sec:construct})};
			
			\node (decoder) [rectangle, minimum width=2.5cm, minimum height=1cm,text centered, text width = 2.5cm, draw=black, fill=blue!25, right of = embedding] {Fully-connected predictor \\ (Section \ref{ss:complete-approach})};
			
			\node (output) [rectangle, rounded corners, minimum width=2.5cm, minimum height=1cm,text centered, text width = 2.5cm, draw=black, fill=green!30, right of = decoder] {Related task\\ (Section \ref{ss:fine-tune-data})};
			
			\node (glm) [rectangle, rounded corners, minimum width=2.5cm, minimum height=1cm,text centered, text width = 2.5cm, draw=black, fill=blue!25, below of = decoder, yshift = 1cm] {GLM \\ (Section \ref{sec:application})};
			
			\node (insurance) [rectangle, rounded corners, minimum width=2.5cm, minimum height=1cm,text centered, text width = 2.5cm, draw=black, fill=green!30, below of = output, yshift = 1cm] {Claims data\\ (Section \ref{ss:insurance-data})};
			
			\node (customer-image) [rectangle, rounded corners, minimum width=2.5cm, minimum height=1cm,text centered, text width = 2.5cm, draw=black, fill=green!35, below of = data, yshift = 1cm] {Customer house image};
			
			\node (representation) [rectangle, minimum width= 6.25cm, minimum height=1cm,text centered, text width = 5.25cm, draw=black, fill=red!25, below of = encoder1, xshift = 1.75cm, yshift = 1cm] {Representation model};
			
			\node (embedding2) [rectangle, rounded corners, minimum width=2.5cm, minimum height=1cm,text centered, text width = 2.5cm, draw=black, fill=violet!35, below of = embedding, yshift = 1cm] {Customer image embedding};
			
			\node (insurance-data) [rectangle, rounded corners, minimum width=2.5cm, minimum height=1cm,text centered, text width = 2.5cm, draw=black, fill=green!35, below of = customer-image, yshift = 1.5cm] {Other variables\\ (Section \ref{ss:insurance-data})};
			
			\node[draw,thick,fit=(encoder1) (encoder2),inner sep=3mm] (representation-box) {};
			
			\draw [->, thick] (data) -- (encoder1);
			\draw [<->, thick] (encoder1) -- (encoder2);
			\draw [<->, thick] (encoder2) -- (embedding);
			\draw [<->, thick] (embedding) -- (decoder);
			\draw [<->, thick] (decoder) -- (output);
			\draw [->, line width=2mm, color = red!25] (representation-box) -- (representation);
			
			\draw [->, thick] (customer-image) -- (representation);
			\draw [->, thick] (representation) -- (embedding2);
			\draw [->, thick] (embedding2) -- (glm);
			
			\draw [<->, thick] (glm) -- (insurance);
			\draw [->, thick] (insurance-data) -| (glm);
	\end{tikzpicture}}
	\caption{Framework for the representation-learning framework.}\label{fig:framework}
\end{figure}

We summarize the framework used for our experiments in Figure \ref{fig:framework}. It is a special case of Figure \ref{fig:unsupervised} but replacing the generic emerging variables with actual components we use for image data. The representation learning framework appears in the first row, while the actuarial pricing model appears in the second and third rows. The main task on the first row is to train a model that inputs the SVI of a house and outputs predictions on insurance-related tasks. One characteristic of this first model is a bottleneck near the end of the model, called image embedding, such that all the information about the input image is condensed into a small vector that will eventually yield a prediction for the related tasks. Because of the bottleneck, the neural network is implicitly regularized such that it is forced to contain all useful information about the related task and little information about other unrelated tasks. Note that most arrows flow both ways in the first row. The forward direction is straightforward since this model takes a house image as input and a prediction of the related tasks as output. The backward direction means that the parameters of the components (image model, dimension reduction and fully-connected predictor) depend on the related task data since one optimizes the parameters of these models to perform well on the related tasks. 

Once the representation-learning step in the first row is completed, we no longer change the parameters of the image model or the dimension reduction models: within our approach, they are considered fixed for the remainder of the framework. We use the trained representation learning model to construct an embedding of the SVI for a potential customer. We then use the customer house image embedding, as well as other variables (traditional actuarial variables or other sources of novel information such as geographic embeddings or textual embeddings) to construct a predictive model for claim count or claim severity. If we construct useful embeddings in the representation-learning step, it is unnecessary to use flexible models for the claim count or claim severity model since all useful non-linear transformations and interactions between parts of the SVI are already captured in the representation-learning step. Note that there are only two-way arrows between the GLM and the claims data since the representation model does not depend on the claim data. It follows that the parameters of the representation model do not contribute to the degrees of freedom in the GLM, so we get the advantages of neural networks (much flexibility to learn useful non-linear transformations and interactions between variables) while still being able to rely on the statistical properties of GLMs through maximum likelihood estimation. 

One notices from Figure \ref{fig:framework} that the house images for training the representation model do not need to be the same as those for insurance pricing. One could train the representation model using data from a city/state/country and use that same model to construct customer image embeddings for houses in another territory. Assuming that there are no significant changes in how the SVI is collected and assuming the training data has seen a wide variety of house styles and years (mix of high and low-income neighbourhoods, a mix of houses/condos/apartments), one will be able to generate predictions on new houses that were not used in constructing the representation model. 

\section{Image data}\label{sec:data}

Most of the applications of images we have encountered in the actuarial science or risk management literature have been dedicated to comparing images before and after an event to examine the status of a property (the occurrence or extent of damages); see Section \ref{ss:related}. What sets our approach apart is that we only use ``before'' images, that is, an image of the insured property in its normal condition. To the best of our knowledge, we are the first to use images as inputs to ratemaking models. For this reason, we must define the desirable attributes for image data and determine which data cleaning is necessary for ratemaking. In this section, we discuss the data we will use to construct embeddings of house images. 

If one uses images as input to a ratemaking model, then the images and their content should be convenient to use and satisfy the properties of rating variables. One place to look for such properties are the Actuarial Standards of Practice (ASOP) from the Actuarial Standards Board (\href{http://www.actuarialstandardsboard.org/}{http://www.actuarialstandardsboard.org/}). Such standards are guidelines on the applications, methods, procedures and techniques actuaries follow when conducting their professional work in the United States. Relevant ASOPs are No. 12 (Risk Classification), No. 23 (Data Quality) and No. 56 (Modeling). Below, we list some considerations that, in our opinion, the image data should have such that they are convenient to use and respect the ASOPs: 
\begin{enumerate}
	\item Relevance: The image should contain relevant information about the house that could affect insurance rates (irrelevant images could lead to irrelevant rates).
	\item Causality: The image should encompass elements that could cause or be impacted by losses, ensuring that the image model can capture the components that generate risk.
	\item Data quality: The images adequately identify risk factors (for instance, high enough resolution).
	\item Representation: The image dataset should contain diverse images and be representative of the population we intend to insure (different types of houses/buildings, different architectural types from many historical periods). They should also be photographed under different weather conditions. Further, one should avoid using a dataset that over or under-represents certain regions. Note that bias could appear in the data if, for instance, the camera quality is higher in wealthier neighbourhoods. 
	\item Availability: Images are available or quickly obtainable for every (most) potential customer for the desired market.
	\item Privacy: The model should not use personally identifiable information or sensitive details that could identify people in the image (for instance, faces of pedestrians, street address, licence plate number). 
	\item Legal compliance: For example, the images should not contain information that reveals protected attributes such as the gender or race of policyholders. If permission to use the image data is required, the insurance company should obtain this permission before beginning the quoting process.
	\item Ground truth data: One has access to a dataset of features describing the images such that we may fine-tune the model (only if using the representation-learning approach).  
\end{enumerate}

Within the context of this paper, the input data will be SVI, and we henceforth use input data, image data and SVI interchangeably. While many sources of image data were considered to answer our research question (see the discussion for more details), we decided to use data from Google Street View for a few reasons. First, images of many houses are available online; hence there is no need for customers to provide images of their homes during the quoting process. Second, they have an easy-to-use API that can easily be incorporated within a quoting process. Therefore, the availability consideration is satisfied. We note that one can apply the machine learning model, initially trained using data from one specific geographical area, to assess or make predictions about a different geographical area as long as the image dataset satisfies the representation consideration across both places. 

\begin{figure}[ht]
	\centering
	\includegraphics[width=0.5\textwidth]{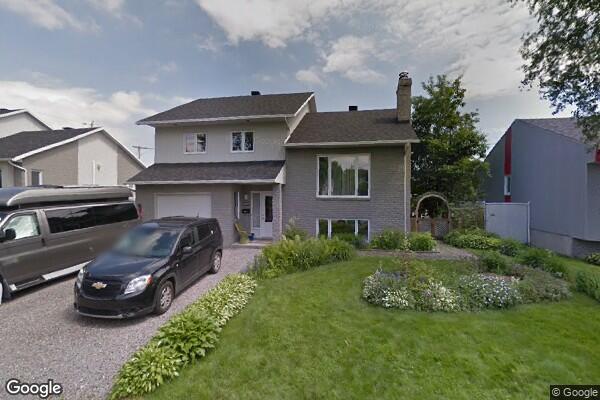}
	\caption{Ideal candidate for facade image.}\label{fig:ideal}
\end{figure}

We present, in Figure \ref{fig:ideal}, an example of the ideal candidate for images that we wish to include within our study. The house in the image is unobstructed, such that information about, for instance, the roofing quality, the facade material, the number of stories or the presence of a garage. Further, the house is centred within the image and contains contextual information such as parts of neighbours' homes, the presence or absence of trees or electrical lines, etc. Finally, the image does not contain sensitive information that an insurance company does not want to consider within the quoting process. This image, therefore, satisfies the relevance, causality, data quality and privacy attributes. 

It is not always true that SVI satisfies most of the desirable properties as in Figure \ref{fig:ideal}. For this reason, one must arrange the dataset before using it for ratemaking purposes. Further, we do not want to do this manually (since we would not want to do this manually within a quoting model in production); we will instead rely on machine learning techniques. We distinguish between two types of cleaning.
\begin{figure}[ht]
	\centering
	\includegraphics[width=0.4\textwidth]{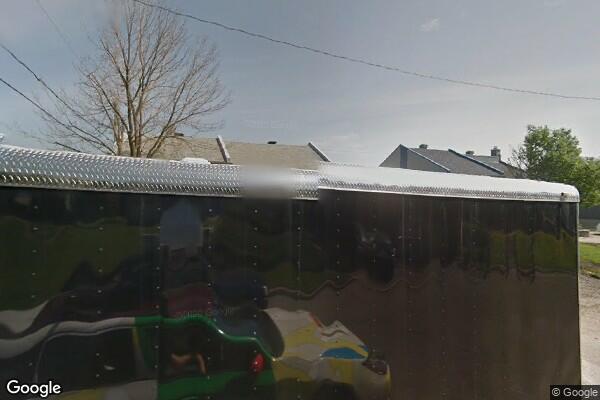} \quad
	\includegraphics[width=0.4\textwidth]{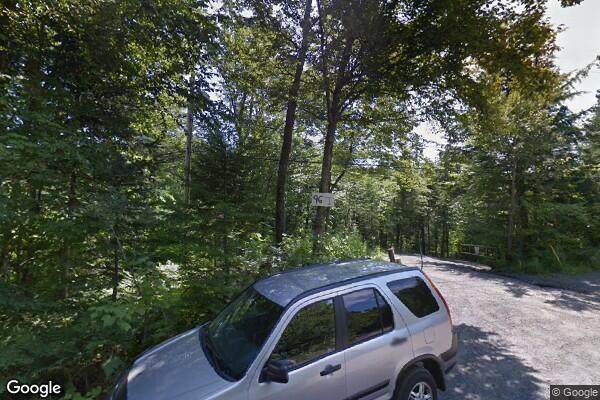}
	\caption{Examples of images requiring filtering.}\label{fig:filtering-example}
\end{figure}
The first is unsalvageable data, that is, containing no information about the dwelling; in that case, we will need to filter out that data and discard that observation since the image does not satisfy the relevance or causality criteria. We present examples of such scenarios in Figure \ref{fig:filtering-example}. These situations happen when an object hides the house of interest (for instance, when a truck is parked in front of the house and completely obstructs the house) or when the house is on a large piece of land (in which case, we may only see the entrance of the property and not the house itself). Other examples are that there is no house in the picture, the image is not available in Google Street View, and the homeowner requested Google to remove the house from Street View (in which case the house will be blurred). 

\begin{figure}[ht]
	\centering	
	\includegraphics[width=0.4\textwidth]{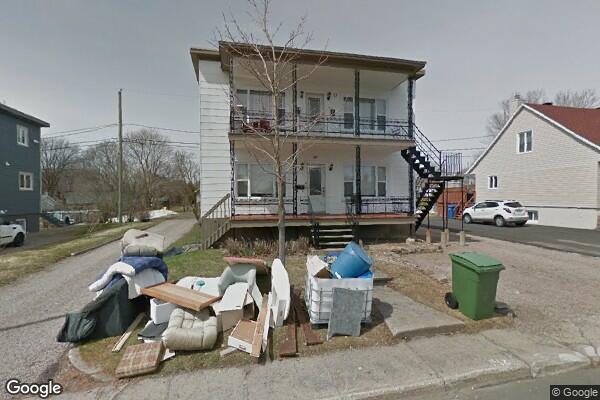}
	\qquad
	\includegraphics[width=0.4\textwidth]{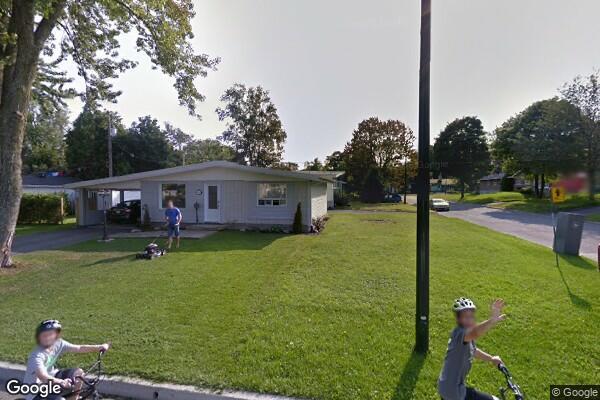}
	\caption{Examples of images requiring censoring.}\label{fig:censoring-example}
\end{figure}

The second is data which may contain information that should not be used for insurance pricing, which may fail the privacy and legal compliance attributes. In these cases, we could remove the segments of the image which cause problems. For instance, items in front of the house temporarily (waste bins, debris, etc.) should not be considered within the image since they do not cause or are not impacted by insurance risks. Such examples are provided in Figure \ref{fig:censoring-example}. 

\begin{figure}[ht]
	\centering	
	\includegraphics[width=0.18\textwidth]{img/10.jpg}
	\includegraphics[width=0.18\textwidth]{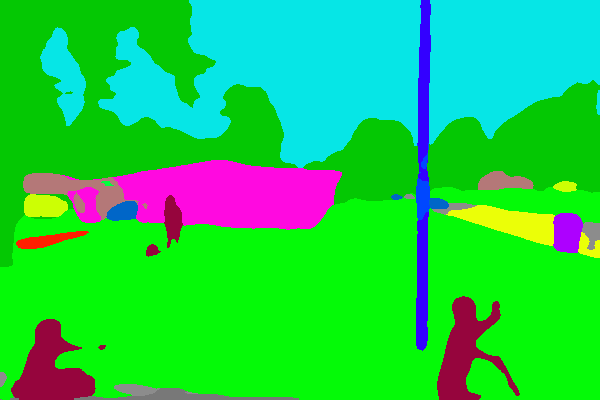}
	\includegraphics[width=0.18\textwidth]{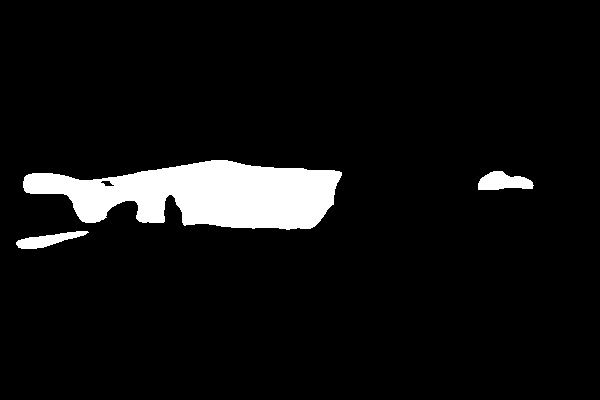}
	\includegraphics[width=0.18\textwidth]{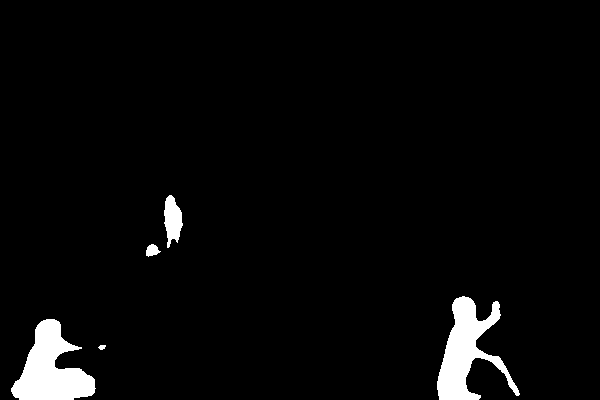}
	\includegraphics[width=0.18\textwidth]{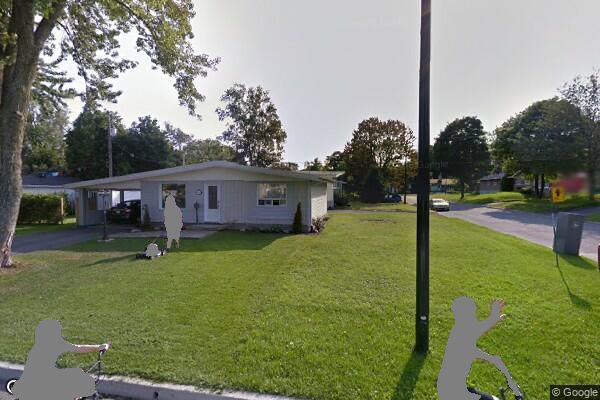}
	\caption{Steps in image cleanup.}\label{fig:censoring-process}
\end{figure}

To identify the cases where we must clean or censor the image, we rely on image segmentation methods. Following \cite{blanc2022caracterisation} and \cite{blierwong2021rethinking}, we use the pre-trained semantic segmentation models from \cite{zhou2017scene} and \cite{zhou2019semantic} with the \texttt{ResNet50dilated + PPM\_deepsup} models to obtain the categories of objects in the images. We present an example image in Figure \ref{fig:censoring-process} with its semantic segmentation (first and second panes). One may compute the percentage of the image that is a house, building or edifice from the mask of these categories (third pane of Figure \ref{fig:censoring-process}). Within our experiment, if less than 5\% of the image is of these categories, we will assume that there is no information about the house to be useful within our application and discard this image. Doing this step removes 4.134\% of the images in our dataset, which means that in practice, the insurance company could use this ratemaking framework for over 95\% of citizens in the city of Québec. 

To mask humans and potentially temporary objects, we obtain the segmentation from the images, then construct a mask for the categories corresponding to person/individual and to a list of objects potentially destined for garbage collection such as seat/desk/lamp/toy/pillow (fourth pane of Figure \ref{fig:censoring-process}). We then replace these pixels with the average pixel value to hide the problematic parts of the image (fifth pane of Figure \ref{fig:censoring-process}).

\section{Representation of related tasks from street view imagery}\label{sec:construct}

In this section, we will apply the framework outlined in the first row of Figure \ref{fig:framework} to the dataset of SVI presented in Section \ref{sec:data}. We will first present the data we selected for the related tasks, that is, the data used to construct the representation learning model. We then present the architecture for the representation models and explain how to extract the embedding vector from the representation learning model. Finally, we will present the training strategy and some information extraction results. 

\subsection{Constructing representations}\label{ss:transfer}

\subsubsection{Related task data}\label{ss:fine-tune-data}

Most readily available image models are pre-trained on ImageNet \cite{deng2009imagenet}, a widely-used dataset containing labelled images that serve as the foundation for training deep neural networks in object recognition, containing over 20,000 classes such as banana, banjo and baseball. Starting with pre-trained image models to train our image model is useful since the early layers of these models learn convolutional filters that identify different patterns in the image. However, the feature map they provide may not be useful for insurance applications. For this reason, we must fine-tune the model to a feature space which may be more related to houses since this is the insured good we are considering within our insurance application. 

One may use many types of information to construct embeddings. However, choosing a related task that is close or similar to our eventual task of interest (predicting future claim frequency or severity) will yield embeddings that are more useful for that task of interest. Another consideration is the availability of the data for the related task. Ideally, the data will have less irreducible uncertainty than insurance loss data and have a high number of observations such that the representation model can learn flexible features. Another aspect of the related task is that there are many of them, such that the intermediate representation does not overfit on one particular aspect of the image but on the general contents of the image that may be related to insurance losses. Within the context of our representation-learning framework, the actuarial priority lies in selecting the appropriate related tasks. One needs a good understanding of risk factors and the insured product to select relevant tasks that yield useful transferable representations. 

Within this paper, we use property assessment data, which is public data collected by the city, to determine the property taxes for a building. These assessments are conducted by an evaluator, whose roles are to keep an inventory of buildings, establish the value of these buildings and justify their opinions. Such opinions are based on the territory, the dimensions of the land, the age of the building (adjusted to account for major renovations or additions), the quality of the construction and the components (materials) as well as the area of the buildings. As such, they provide information about the house, and fine-tuning a model on this data may produce a feature space that is more useful for home insurance pricing. 

In our implementation, we use property tax data from the city of Québec in the province of Québec, Canada. We chose this dataset because it was readily available online and contained information for all 182,419 images within the property assessment dataset, with values corresponding to a market date of July 1, 2018. Table \ref{tab:summary-assessment} summarizes the structured data available in this dataset. We only considered a subset of variables, including the number of floors, construction year, land value, building value, and total value. The reason behind this subset was the substantial amount of missing values in the other variables, which hindered precise modelling. We dropped any observations with missing values, which resulted in a training dataset size of 172,098. However, variables like the number of units and area were challenging to predict accurately due to the limitations of the data. For example, buildings with many units may be too tall to capture the entire structure in an image, while land area is difficult to predict using only a facade image and not a full property image.
\begin{table}[ht]
	\centering
	\resizebox{\textwidth}{!}{
	\begin{tabular}{lrrrrr}
		Feature               & Minimum &  Average &     Maximum & Standard deviation & \% missing \\ \hline
		Number of floors      &       1 &     1.55 &          33 &               0.86 &       3.51 \\
		Construction year     &    1604 &     1971 &        2018 &              30.15 &       8.52 \\
		Land value            &       1 &  238,387 & 175,781,000 &          1,226,238 &       0.15 \\
		Building value        &       1 &  435,890 & 652,421,000 &          5,047,336 &          0 \\
		Total value           &       1 &  673,896 & 828,202,000 &          6,123,145 &          0 \\
		Units                 &       1 &     3.59 &         743 &              19.22 &       7.14 \\
		Non-residential units &       1 &     2.89 &         416 &               8.92 &      92.81 \\
		Rental units          &       1 &    34.66 &         611 &              60.89 &      99.54 \\
		Front measure $(m)$   &    0.12 &    23.47 &     118,367 &             566.26 &      32.72 \\
		Area $(m^2)$          &    0.10 & 1,800.89 &   6,759,800 &          26,199.23 &       3.05 \\ \hline
	\end{tabular}}
	\caption{Variables and summary statistics from the property assessment dataset.}\label{tab:summary-assessment}
\end{table}
To limit the impact of outliers, we have applied several transformations to our dataset. Rather than using the construction year, we use the age of the building (with respect to the evaluation date of 2018). We have also capped the building age at 100 to ensure consistency and reduce the influence of outliers. Additionally, we have applied a logarithmic transformation to the land, building, and total values to help normalize their distributions. However, we have removed five observations where the land, building, or total value was listed as 1\$, as these appeared to represent empty lots and may have skewed the analysis.

\subsubsection{General structure of the representation models for computer vision}

Models for computer vision are designed to analyze data such as images. These models typically use convolutional neural networks, a neural network suited to processing data with local information \cite{alzubaidi2021review}. Image models based on deep learning have revolutionized computer vision by achieving state-of-the-art performance on many image-related tasks, such as image classification, image segmentation and object detection. These models are trained on large datasets of images and can automatically identify patterns relevant to a certain task. For an introduction to convolutional neural networks in actuarial science, see the supplementary materials of \cite{blierwong2022geographic} or Chapter 9 of \cite{wuthrich2023statistical}.

The general structure of convolutional neural networks (CNNs) has remained the same since the earliest CNN models (see, for instance, \cite{lecun1998gradient}). A typical image model contains two parts with trainable parameters; see Section 14.3 of \cite{murphy2022probabilistica} for a review of popular architectures. The first is a set of convolutional layers which take as input an image and apply convolutional filters along the images. We call this the \textit{Image model} in Figure \ref{fig:framework}. Such convolutional operations identify different patterns and shapes and, with enough depth, may construct a feature map that captures the information inside the image. If the image dataset contains diverse objects, then the feature map of the image will contain very general representations. We will refer to the first part of the image model as the convolutional part and the output of the first part as the feature map from the images. An intermediate part will flatten or unroll the feature map from the images (typically three-dimensional) into a single vector. The second part is a fully-connected set of neural network layers (usually only one) that will use the feature map from the images as input and a predictive task as output. For classification tasks, the output is one value for each category; for multitask regression, the output is one value for each regression task. 

The fully-connected layer between the last hidden layer and the prediction has the same structure as a (generalized) linear model. Therefore, we can interpret the final hidden layer as a condensed representation that contains all of the useful information about the house image to predict the response variables in the related tasks. For this reason, we will consider the final hidden layer of the fully-connected set of layers to be the embedding associated with the house image. To recap, the fully-connected part of the representation framework has a set of layers devoted to simultaneously reducing the dimension of the feature space and capturing non-linear transformations and interactions between the feature maps from the images, which we call \textit{Dimension reduction} in Figure \ref{fig:framework}. The output of this first part of fully-connected layers is the last hidden layer, which contains the representation of the house image, called \textit{Image embedding} in Figure \ref{fig:framework}. The final fully-connected layer performs the role of a (generalized) linear model to predict the related tasks from the embeddings; we call this part the \textit{Fully-connected predictor} in Figure \ref{fig:framework}.

\subsubsection{Complete approach: fine-tuned image model}\label{ss:complete-approach}

To make sure that the embeddings are useful for our eventual task of interest (predicting claim frequency/severity), we must ``guide'' our embeddings toward a feature space that is useful for insurance-related tasks, which implies that we must adapt the weights of the image model and fully-connected layers towards an attractive embedding space. To do this, we start with a pre-trained model that contains representations that are useful for general image classification. That is, the architecture of the set of convolutional layers remains the same. Then, we remove the original set of fully-connected layers and replace them with our own to control the size of the feature maps. The architecture of our fully-connected layers goes from the image model feature space size down to 128, then down to the embedding size, and finally down to the size of the output task. The representation of the image will be the hidden features in the embedding layer, that is, the last hidden features before the predictions. 

Within the complete approach, we allow the weights of the convolutional neural network, the fully-connected layers in the dimension reduction step and the fully-connected predictor to be trained on the related task. We apply the backpropagation algorithm to every trainable parameter in the first row of Figure \ref{fig:framework}. This model is the most flexible and takes the longest to train. 

\subsubsection{Limited approach: frozen image model}

In the second example of unsupervised transfer learning, we use the same framework as the complete approach but allow less parameter flexibility. We keep the parameters of the convolutional neural network fixed and only train the new fully-connected layers. Within the framework outlined in Figure \ref{fig:framework}, this means that when constructing the model for the related task, one does not change the parameters from the image model (they remain the pre-trained parameters on ImageNet classification tasks) but one may train the parameters from the fully-connected layer in the dimension reduction part and the fully-connected predictor part. This means that feature maps from the images are kept the same as the original pre-trained image models. 

In machine learning jargon, weights kept fixed are said to be frozen; hence we call this model the frozen image representation. The limited approach is equivalent to extracting the flattened feature maps from each image in our dataset and training a fully-connected neural network according to the structure specified in the previous section. Therefore, for a fixed dataset, one may compute the feature map from the image model once, store this feature map and exclusively train the fully-connected neural network. From a training perspective, this makes training much more efficient since one does not have to pass the image in the image model (and apply the backpropagation algorithm to the weights of the image model) for every epoch. Within our implementation (that we will detail in Section \ref{ss:experiments}), over 98\% of the weights are in the image model. 

\begin{figure}[ht]
	\centering
	\resizebox{\textwidth}{!}{
		\begin{tikzpicture}
			
			\tikzstyle{connection}=[ultra thick,every node/.style={sloped,allow upside down},draw=\edgecolor,opacity=0.7]
			\tikzstyle{copyconnection}=[ultra thick,every node/.style={sloped,allow upside down},draw={rgb:blue,4;red,1;green,1;black,3},opacity=0.7]
			
			\coordinate (left) at (7.5em, 0);
			\coordinate (right) at (51em, 0);
			\coordinate (top) at (20em, 8em);
			\coordinate (bottom) at (20em, -5.3em);
			\node[draw,line width=1mm, red, dotted, fit=(left) (right) (top) (bottom)] {};
			
			\coordinate (left) at (32em, 0);
			\coordinate (right) at (50em, 0);
			\coordinate (top) at (40em, 6em);
			\coordinate (bottom) at (40em, -4em);
			\node[draw,line width=1mm, blue, dotted, fit=(left) (right) (top) (bottom)] {};
			
			\pic[shift={(0,0,0)}] at (0,0,0) 
			{Box={
					name=conv1,
					fill=green,
					caption=,
					opacity=0.25,
					height=16,
					width=8,
					depth=16
				}
			};
			
			\pic[shift={( 2.2,0,0)}] at (conv1-east) 
			{Box={
					name=conv2,
					caption= ,
					xlabel={{, }},
					zlabel=,
					fill=red,
					opacity=0.25,
					height=8,
					width=16,
					depth=8
				}
			};
			
			\pic[shift={( 2.2,0,0)}] at (conv2-east) 
			{Box={
					name=fc1,
					caption=,
					xlabel={{, }},
					zlabel=,
					fill=white,
					opacity=0.25,
					height=1,
					width=1,
					depth=48
				}
			};
			
			\pic[shift={( 4,0,0)}] at (fc1-east) 
			{Box={
					name=fc2,
					caption= ,
					xlabel={{, }},
					zlabel=\Large ,
					fill=blue,
					opacity=0.25,
					height=1,
					width=1,
					depth=16
				}
			};
			
			\pic[shift={( 2.2,0,0)}] at (fc2-east) 
			{Box={
					name=soft1,
					caption=,
					xlabel={{,}},
					zlabel=,
					fill=violet,
					opacity=0.3,
					height=1,
					width=1,
					depth=8
				}
			};
			
			\pic[shift={( 2.2,0,0)}] at (soft1-east) 
			{Box={
					name=fc3,
					caption=,
					xlabel={{,}},
					zlabel=,
					fill=blue,
					opacity=0.25,
					height=1,
					width=1,
					depth=16
				}
			};
			
			\pic[shift={( 2.2,0,0)}] at (fc3-east) 
			{Box={
					name=fc4,
					caption=,
					xlabel={{,}},
					zlabel=,
					fill=green,
					opacity=0.25,
					height=1,
					width=1,
					depth=8
				}
			};
			
			\draw[->, shorten <=2pt,shorten >=2pt, thick] (conv1-east) -- (conv2-west);
			\draw[->, shorten <=2pt,shorten >=8pt, thick] (conv2-east) -- (fc1-west);
			\draw[->, shorten <=8pt,shorten >=8pt, thick] (fc1-east) -- (fc2-west);
			\draw[->, shorten <=8pt,shorten >=8pt, thick] (fc2-east) -- (soft1-west);
			\draw[->, shorten <=8pt,shorten >=8pt, thick] (fc2-east) -- (soft1-west);
			\draw[->, shorten <=8pt,shorten >=8pt, thick] (soft1-east) -- (fc3-west);
			\draw[->, shorten <=8pt,shorten >=8pt, thick] (fc3-east) -- (fc4-west);
			
			\node[text width = 6em] at (2em, -8em) {\large House \\ image};
			\node[text width = 5em] at (13em, -8em) {\large Image \\ model};
			\node[text width = 5em] at (22em, -8em) {\large Unroll};
			\node[text width = 5em] at (35em, -8em) {\large Fully-\\connected \\ layers};
			\node[text width = 6em] at (42em, -8em) {\large Embedding \\ vector};
			\node[text width = 5em] at (48em, -8em) {\large Fully-\\connected \\predictor};
			\node[text width = 5em] at (54em, -8em) {\large Related \\ tasks};
			
			\node[text width = 15em, anchor=west, align = center] at (10em, 6em) {\LARGE \textcolor{red}{Complete approach: fine-tuned model}};
			\node[text width = 15em, anchor=west, align = center] at (33.5em, 4em) {\LARGE \textcolor{blue}{Limited approach: frozen model}};
			
	\end{tikzpicture}}
	\caption{Architecture for the complete and limited representation approaches.}\label{fig:complete-approach-graph}
\end{figure}
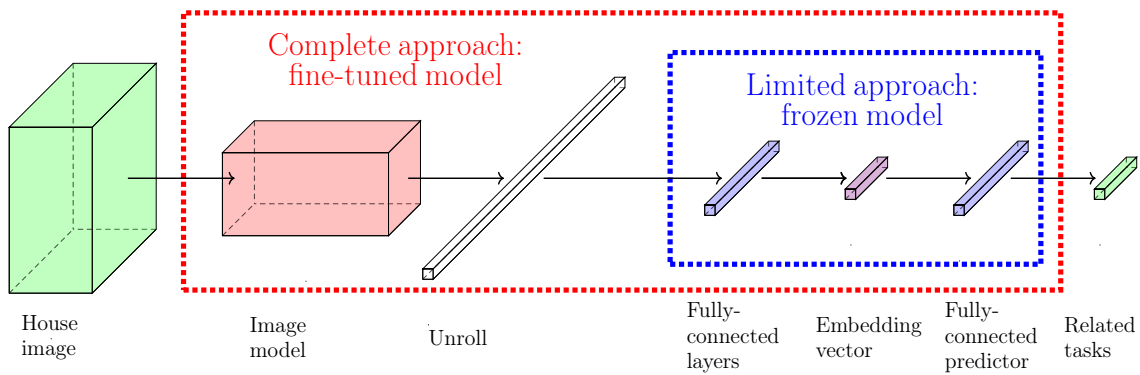

We summarize the fine-tuned and frozen model in Figure \ref{fig:complete-approach-graph}. Within the smaller dotted box lies the fully-connected layers and the fully-connected predictor, whose weights are trainable with respect to the related task. Similarly, within the larger dotted box, the weights from the image model are also trainable with respect to the same tasks. 

\subsubsection{Basic approach: PCA with no fine-tuning}

\begin{figure}[ht]
	\centering
	\resizebox{0.8\textwidth}{!}{
		\begin{tikzpicture}
			
			\tikzstyle{connection}=[ultra thick,every node/.style={sloped,allow upside down},draw=\edgecolor,opacity=0.7]
			\tikzstyle{copyconnection}=[ultra thick,every node/.style={sloped,allow upside down},draw={rgb:blue,4;red,1;green,1;black,3},opacity=0.7]

			\pic[shift={(0,0,0)}] at (0,-8,0) 
			{Box={
					name=conv1,
					fill=green,
					caption=,
					opacity=0.25,
					height=16,
					width=8,
					depth=16
				}
			};
			
			\pic[shift={(2.2,0,0)}] at (conv1-east) 
			{Box={
					name=conv2,
					caption= ,
					xlabel={{, }},
					zlabel=,
					fill=red,
					opacity=0.25,
					height=8,
					width=16,
					depth=8
				}
			};
			
			\pic[shift={( 2.2,0,0)}] at (conv2-east) 
			{Box={
					name=fc1,
					caption=,
					xlabel={{, }},
					zlabel=,
					fill=white,
					opacity=0.25,
					height=1,
					width=1,
					depth=48
				}
			};

			\pic[shift={( 4.5,0,0)}] at (fc1-east) 
			{Box={
					name=soft1,
					caption=,
					xlabel={{,}},
					zlabel=,
					fill=violet,
					opacity=0.3,
					height=1,
					width=1,
					depth=8
				}
			};
			
			\draw[->, shorten <=2pt,shorten >=2pt, thick] (conv1-east) -- (conv2-west);
			\draw[->, shorten <=2pt,shorten >=8pt, thick] (conv2-east) -- (fc1-west);
			\draw[->, shorten <=8pt,shorten >=8pt, thick] (fc1-east) -- (soft1-west);
			
			\node[text width = 6em] at (2em, -29em) {\large House \\ image};
			\node[text width = 5em] at (13em, -29em) {\large Image \\ model};
			\node[text width = 5em] at (22em, -29em) {\large Unroll};
			\node[text width = 12em] at (36em, -29em) {\large Principle components \\ (embedding vector)};

			
			
	\end{tikzpicture}}
	\caption{Architecture for the basic PCA approach.}\label{fig:basic-approach-graph}
\end{figure}
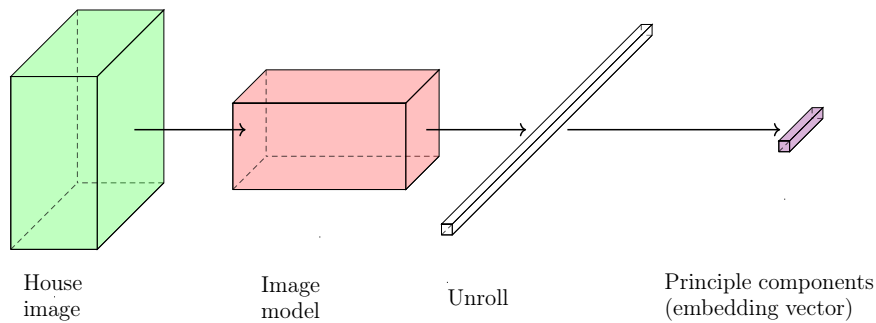

The final approach to constructing embeddings is one where there is no fine-tuning in related tasks. To do this, we start with a pre-trained image model and use the feature map from the images. However, this representation is too large to include within the ratemaking model (512, 1024 or 2048 dimensions), so we apply principal component analysis to reduce the dimension to a more suitable embedding size. We present a diagram of the basic approach in Figure \ref{fig:basic-approach-graph}.

The basic approach, which we call the PCA approach, depends not on the related tasks (notice the absence of related tasks in Figure \ref{fig:basic-approach-graph} compared to Figure \ref{fig:complete-approach-graph}) but on the representations learned on the ImageNet dataset. As such, the representations generated from this approach will capture general notions of shapes and categories but may not be directly appropriate for insurance tasks. However, it requires no fine-tuning tasks and almost no training time, making it a simple model to investigate whether there is useful information within images. We present a summary of the dimension reduction approaches in Table \ref{tab:summary-approaches}; this summary illustrates which parts of Figure \ref{fig:framework} can be trained (flexible) and which maintains their pre-trained weights from ImageNet (frozen). 

\begin{table}[ht]
	\resizebox{\textwidth}{!}{\begin{tabular}{cccc}
			& Image model & Dimension reduction & Fully-connected predictor \\ \hline
			Fine-tuned model     &  Flexible   &      Flexible       &         Flexible          \\
			Frozen model       &   Frozen    &      Flexible       &         Flexible          \\
			PCA with no fine-tuning &   Frozen    &         PCA         &            NA             \\ \hline
	\end{tabular}}
	\caption{Summary of fine-tuned, frozen and PCA construction approaches.}\label{tab:summary-approaches}
\end{table}

\subsection{Experiments}\label{ss:experiments}

\subsubsection{Backbone image models}\label{ss:backbone-image-model}

In this section, we will construct representations of image houses. For representation learning, it is the output of the first part that is of interest to us since this contains general information. We wish to use the embeddings within the representation pricing framework in Figure \ref{fig:framework}. We will consider some of the most popular image models, deep residual neural networks (ResNet, introduced by \cite{he2016deep}) and densely connected convolutional networks (DenseNet, introduced by \cite{huang2017densely}). For the ResNet models, one has pre-trained models for 18, 34, 50, 101 and 152 layers. The 152-layer model does not fit on our GPU, so we did not consider it. We also did not consider the 34-layer model to limit the length of the analysis. Pretrained DenseNet models are available as 121, 169, 201 and 264-layer models, but we only consider the 121-layer model since the larger ones did not fit on our GPU. All four models are pre-trained on the ImageNet dataset \cite{deng2009imagenet}. 

\subsubsection{Training strategy}

We have selected five related tasks to train the representation models: the construction year, the land value (log), the building value (log), and the total value (log), denoted respectively by $y_1,y_2,y_3$ and $y_4$, and the number of floors. For the number of floors, we cap the values at three since the facade image for buildings with many floors will only show the first few floors. We consider the number of floors as a categorical variable (1, 2 or 3+ floors) and use a classification loss for this task; we denote these variables as $y_5, y_6$ and $y_7$. The number of observations for each class is 99398, 59915 and 12785. The remaining tasks are treated as regression. The loss function is an equally-weighted average of the mean squared error loss for the regression tasks and the cross-entropy loss for the classification task. We do not regularize the loss function. Therefore, the function to minimize is 
$$\sum_{i =1}^{3}(y_i - \hat{y}_i)^2 - \sum_{i = 4}^{7} y_i \log \hat{y}_i,$$ 
where $\hat{y}_i$ are the predictions for $y_i$, for $i = 1, \dots, 7$. 

In our conversations with a few insurance companies, we know that typical ratemaking models for home insurance use between 30 and 100 rating variables. For that reason, we will consider embeddings of dimensions 8, 16 and 32 throughout this paper since we do not want to consider larger embedding dimensions such that the SVI features outnumber the traditional features.

To construct representations from unsupervised transfer learning, we follow the same training strategy for every model considered. We start with some pre-trained image models (ResNet and DenseNet) and get rid of the last layer (the one going from a feature space to a classification space). Instead, we add three fully-connected layers after the feature space, which we call the fine-tuning block. The first goes from the feature space size (depending on the model, 512, 1024 or 2048) to 128. Then, from 128 to the embedding size (8, 16 or 32; this acts as a hyperparameter), and the last from the embedding size to the output size (7). Between each layer, we use a LeakyReLU activation with a negative slope of 0.1, that is, $\max(0, x) - 0.1 \min(0, x)$. We initialize each fully-connected layer with Xavier initialization, proposed in \cite{glorot2010understanding}, with the standard ``gain'' parameter of one. The bias for each fully-connected layer is initialized at zero. 

To train the model, we set the batch size as the largest power of two such that the data and model fit on the GPU (GeForce RTX 20 with 8GB of RAM). We train the model for 25 epochs with an initial learning rate of $10^{-4}$. Every five epochs, we reduce the learning rate by a factor of ten. We summarize the models in Table \ref{tab:resnet-parameters}.

\begin{table}[ht]
	\centering
	\begin{tabular}{lrrrr}
		&  ResNet18 &  ResNet50 & ResNet101 & DenseNet121 \\ \hline
		Convolution weights            & 11,176,512 & 23,508,032 & 42,500,160 &    6,953,856 \\
		Image model feature space      &        512 &       2024 &       2024 &         1024 \\
		FC weights (8)                 &     65,664 &    263,304 &    263,304 &      132,232 \\
		FC weights (16)                &     67,728 &    264,336 &    264,336 &      133,264 \\
		FC weights (32)                &     69,792 &    266,400 &    266,400 &      135,328 \\
		Batch size                     &        128 &         32 &         32 &           32 \\ \hline
	\end{tabular}  
	\caption{Summary of parameters for ResNet and DenseNet models.}\label{tab:resnet-parameters}
\end{table}

\subsubsection{Experimental results}

To answer our research question, we only require the intermediate representations to be related to insurance. For this reason, we do not need to (i) regularize the network, (ii) perform out-of-sample model validation, (iii) calibrate the predictions or (iv) find the best model or the best training strategy possible. Therefore, we only present results for in-sample training. Readers interested in information extraction from facade images should refer to \cite{blanc2022caracterisation}. 

\begin{table}[ht]
	\centering
	\begin{tabular}{cccccc}
		&               & ResNet-18 & ResNet-50 & ResNet-101 & DenseNet-121 \\ \hline
		\multirow{4}{*}{\makecell{Frozen\\ image model}}   &   Loss (8)    &   1.261   &   1.281   &   1.252    &    1.184     \\
		&   Loss (16)   &   1.267   &   1.199   &   1.197    &    1.162     \\
		&   Loss (32)   &   1.172   &   1.161   &   1.143    &    1.125     \\
		& Training time &  64 min   &  171 min  &  166 min   &    97 min    \\ \hline
		\multirow{4}{*}{\makecell{Fine-tuned\\ image model}} &   Loss (8)    &   0.078   &   0.075   &   0.051    &    0.098     \\
		&   Loss (16)   &   0.055   &   0.069   &   0.051    &    0.082     \\
		&   Loss (32)   &   0.055   &   0.069   &   0.050    &    0.077     \\
		& Training time &  981 min  & 3779 min  &  5583 min  &   5076 min   \\ \hline
	\end{tabular}  
	\caption{Summary of results of training for Image models.}\label{tab:resnet-frozen-results}
\end{table}

\def\myConfMatFrozen{{
		{81.83, 21.14, 16.34}, 
		{16.14, 71.38, 22.19}, 
		{2.04,  7.48, 61.47}, 
}}
\def\myConfMatFT{{
		{99.8,  0.26,  0.04}, 
		{0.18, 99.72,  0.09}, 
		{0.03,  0.03, 99.86}, 
}}
\def\classNames{{"1","2","3+"}} 
\def\numClasses{3}
\def\myScale{1.5}

In Table \ref{tab:resnet-frozen-results}, we summarize the loss values for the different models considered. Let us offer a few remarks on the performance of the information extraction models. First, the training time remained stable no matter the embedding size since they had approximately the same number of parameters (see Table \ref{tab:resnet-parameters}). Training all models takes over 33 days on a GeForce RTX 20 with 8GB of RAM. For both the frozen image model and the fine-tuned image model, the best architecture is given by the ResNet-101 model with 32 embeddings, corresponding to the model with the most parameters. In general, increasing the size of the embedding layer leads to a lower loss, which makes sense since the models are more flexible. The only exception is for the frozen image model with ResNet-18, where the loss for embedding size 8 is lower than that for embedding size 16, which could be due to different local minima (training the same models with different initial weights should recover the correct ordering). For the fine-tuned image model, the difference in loss functions is very small, meaning that 16 embedding dimensions may be sufficient to learn the predictive tasks and that adding 16 more leads to redundant information. The DenseNet-121 model performed the best on the Frozen image model but worse on the fine-tuned image model. One possible explanation is that the DenseNet representations were more useful (which explains why they perform well on the image model) but that the higher number of parameters in the ResNet models enabled them to perform better when every parameter in the neural network was allowed to vary. Also, from Table \ref{ss:transfer}, one observes that the percentage of variance captured by the first principal components of the feature space generated by the image models is larger for ResNet models compared with the DenseNet models, meaning that the feature space of DenseNet models are more linearly independent, meaning that DenseNet captures more non-linear effects for the same embedding dimension. 

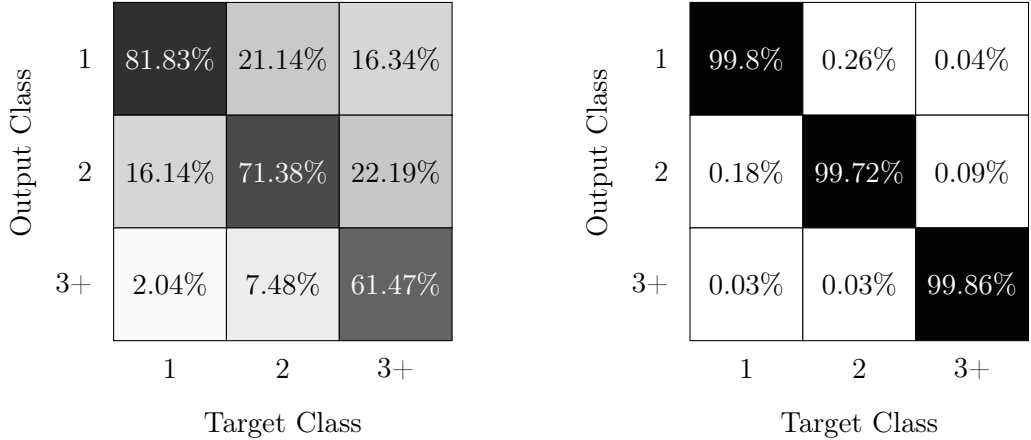
\begin{figure}
	\centering
	\begin{tikzpicture}[
		scale = \myScale,
		]
		
		\tikzset{vertical label/.style={rotate=90,anchor=east}}
		\tikzset{diagonal label/.style={rotate=45,anchor=north east}}
		
		\foreach \y in {1,...,\numClasses}
		{
			\node [anchor=east] at (0.4,-\y) {\pgfmathparse{\classNames[\y-1]}\pgfmathresult}; 
			
			\foreach \x in {1,...,\numClasses}  
			{
				\def\totSamples{0}
				\foreach \ll in {1,...,\numClasses}
				{
					\pgfmathparse{\myConfMatFrozen[\ll-1][\x-1]}   
					\xdef\totSamples{\totSamples+\pgfmathresult} 
				}
				\pgfmathparse{\totSamples} \xdef\totSamples{\pgfmathresult}  ----------------
				
				\begin{scope}[shift={(\x,-\y)}]
					\def\mVal{\myConfMatFrozen[\y-1][\x-1]}
					\pgfmathsetmacro{\r}{\mVal} 
					\pgfmathtruncatemacro{\p}{\r/\totSamples*100}
					\coordinate (C) at (0,0);
					\ifthenelse{\p<50}{\def\txtcol{black}}{\def\txtcol{white}} 
					\node[
					draw,
					text=\txtcol,
					align=center, 
					fill=black!\p, 
					minimum size=\myScale*10mm,
					inner sep=0, 
					] (C) {\r\%};
					\ifthenelse{\y=\numClasses}{
						\node [] at ($(C)-(0,0.75)$)
						{\pgfmathparse{\classNames[\x-1]}\pgfmathresult};}{}
				\end{scope}
			}
		}
		
		\coordinate (yaxis) at (-0.3,0.5-\numClasses/2);
		\coordinate (xaxis) at (0.5+\numClasses/2, -\numClasses-1.25);
		\node [vertical label] at (yaxis) {Output Class};
		\node []               at (xaxis) {Target Class};
	\end{tikzpicture}
	\qquad\qquad
	\begin{tikzpicture}[
		scale = \myScale,
		]
		
		\tikzset{vertical label/.style={rotate=90,anchor=east}}
		\tikzset{diagonal label/.style={rotate=45,anchor=north east}}
		
		\foreach \y in {1,...,\numClasses}
		{
			\node [anchor=east] at (0.4,-\y) {\pgfmathparse{\classNames[\y-1]}\pgfmathresult}; 
			
			\foreach \x in {1,...,\numClasses}  
			{
				\def\totSamples{0}
				\foreach \ll in {1,...,\numClasses}
				{
					\pgfmathparse{\myConfMatFT[\ll-1][\x-1]}   
					\xdef\totSamples{\totSamples+\pgfmathresult} 
				}
				\pgfmathparse{\totSamples} \xdef\totSamples{\pgfmathresult}  ----------------
				
				\begin{scope}[shift={(\x,-\y)}]
					\def\mVal{\myConfMatFT[\y-1][\x-1]}
					\pgfmathsetmacro{\r}{\mVal} 
					\pgfmathtruncatemacro{\p}{\r/\totSamples*100}
					\coordinate (C) at (0,0);
					\ifthenelse{\p<50}{\def\txtcol{black}}{\def\txtcol{white}} 
					\node[
					draw,
					text=\txtcol,
					align=center, 
					fill=black!\p, 
					minimum size=\myScale*10mm,
					inner sep=0, 
					] (C) {\r\%};
					\ifthenelse{\y=\numClasses}{
						\node [] at ($(C)-(0,0.75)$)
						{\pgfmathparse{\classNames[\x-1]}\pgfmathresult};}{}
				\end{scope}
			}
		}
		
		\coordinate (yaxis) at (-0.3,0.5-\numClasses/2);
		\coordinate (xaxis) at (0.5+\numClasses/2, -\numClasses-1.25);
		\node [vertical label] at (yaxis) {Output Class};
		\node []               at (xaxis) {Target Class};
	\end{tikzpicture}
	\caption{Confusion matrices of the number of stories for frozen (left) and fine-tuned (right) models with ResNet-101 and 32 embedding dimensions.}
	\label{fig:confusion}
\end{figure}

We now provide prediction results for the ResNet-101 model with 32 embedding dimensions since it has the lowest loss value. In Figure \ref{fig:confusion}, we present the confusion matrix for the number of stories in the Frozen and Fine-tuned image models. The root mean squared error (RMSE) for the regression tasks is in Table \ref{tab:rmse-regression}. Recall that these values are for training datasets; we do not perform out-of-sample prediction since our primary goal is to construct useful representations. One notices from the confusion matrix that both models identify the correct class for most cases. Further, if a model misclassifies the number of floors, it is more likely to over/underestimate by one floor rather than by two floors. From Table \ref{tab:rmse-regression}, one observes that the RMSE for the regression tasks is quite high. One reason may be that age is a difficult category to predict since an old house could be renovated to look like new; hence it is reasonable for a predictive model to predict a low age for a centenarian renovated house, yielding high RMSE values. Further, land value is highly dependent on the land size and the land location. While the image may contain information about the frontal measure of the land, it does not have access to information about the location. It may be this reason that land value is the worst regression task out of the three monetary variables. The building value is a proxy of the size and quality of the home, which is more useful for insurance contexts. For that variable, the fine-tuned model had a better predictive performance. 

\begin{table}[ht]
	\centering
	\begin{tabular}{ccccc}
		&  Age  & Total value (log) & Building value (log) & Land value (log) \\ \hline
		Frozen   & 25.91 &      0.5732       &        0.5422        &      0.7343      \\
		Fine-tuned & 31.22 &      0.6580       &        0.3929        &      0.7703      \\ \hline
	\end{tabular}
	\caption{Root mean squared error of regression tasks for frozen and fine-tuned models with ResNet-101 and 32 embedding dimensions.}\label{tab:rmse-regression}
\end{table}

Let us finally look at the performance of PCA on the feature space from the images. In Figure \ref{fig:percentage-var-explained}, we present the ranked percentage of variance explained, while in Table \ref{tab:pca-variance}, we present the cumulative variance explained by the first few principal components. Note that while the percentage of variance captured by all models seems to follow a similar pattern, the feature space for different models is different (see Table \ref{tab:pca-variance} for an overview, in particular, the feature space for ResNet50 is four times larger than the feature size for ResNet18). 
\begin{figure}[ht]
	\centering
	\resizebox{0.8\textwidth}{!}{
		\begin{tikzpicture}
			\begin{axis}[
				width = 5in, 
				height = 2.5in,
				ymin = 0,
				xmin = 0, 
				ymax = 0.06, 
				xmax = 64,
				xlabel={Number of principal components},
				xtick={2, 4, 8, 16, 32, 64, 128},
				ylabel={Percentage of variance explained}, 
				legend style={at={(1,1)},anchor=north east},
				]
				
				\addplot[red] table [x=x, y=var, col sep=comma] {data/embeddings-resnet-18-pca-cumul-variance.csv};
				\addlegendentry{ResNet18}
				\addplot[orange] table [x=x, y=var, col sep=comma] {data/embeddings-resnet-50-pca-cumul-variance.csv};
				\addlegendentry{ResNet50}
				\addplot[green] table [x=x, y=var, col sep=comma] {data/embeddings-resnet-101-pca-cumul-variance.csv};
				\addlegendentry{ResNet101}
				\addplot[blue] table [x=x, y=var, col sep=comma] {data/embeddings-densenet-121-pca-cumul-variance.csv};
				\addlegendentry{DenseNet121}
				\addplot[dotted, samples=5, smooth] coordinates {(8,0)(8,0.06)};
				\addplot[dotted, samples=5, smooth] coordinates {(16,0)(16,0.06)};
				\addplot[dotted, samples=5, smooth] coordinates {(32,0)(32,0.06)};
			\end{axis}
		\end{tikzpicture}
	}
	\caption{Percentage of variance explained for the first principal components for the feature spaces.}\label{fig:percentage-var-explained}
\end{figure}
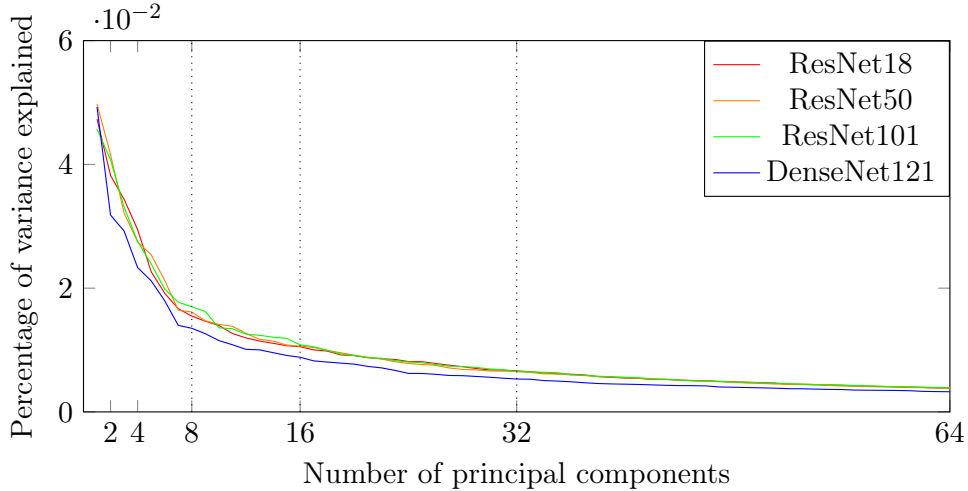
\begin{table}[ht]
	\centering
	\begin{tabular}{lrrrrr}
		Model            & Feature space size &    8 &   16 &   32 &   64 \\ \hline
		ResNet18-PCA     &                512 & 0.22 & 0.32 & 0.45 & 0.61 \\
		ResNet50-PCA     &               2048 & 0.23 & 0.33 & 0.46 & 0.61 \\
		ResNet101-PCA    &               2048 & 0.23 & 0.33 & 0.46 & 0.62 \\
		DenseNet121-PCA  &               1024 & 0.20 & 0.28 & 0.39 & 0.52 \\ \hline
	\end{tabular}
	\caption{Cumulative percentage of variance captured by the first principal components.}\label{tab:pca-variance}
\end{table}

\section{Actuarial application}\label{sec:application}

Let us now summarise what we have accomplished and what is yet to come. We have collected an SVI dataset and a municipal evaluation dataset and cleaned out protected or temporary attributes from the images. Through three training strategies (PCA, frozen and fine-tuned), we have constructed embeddings of image houses. For the frozen and fine-tuned models, we trained these embeddings with the hope that they will be useful for insurance-related tasks such as predicting the number of stories in the house, the construction year and the building/land/total value of the property. As such, the representations will focus on the house and ignore other information which might not be related. The embeddings, therefore, act as proxies for the notion of the size of the house, the age or quality of the house and the value of the house. In this section, we will use embeddings as inputs to a ratemaking model. It is important to mention that for the entire actuarial application, the embeddings remain fixed and cannot be changed. Moreover, this marks our initial use of the insurance dataset, which implies that the claims data did not influence the construction of the embeddings. We perform the extrinsic evaluation of our embeddings on the insurance dataset. To do so, we follow the framework from \cite{blierwong2021rethinking, blierwong2022geographic} and train a GLM using only the embeddings. 

\subsection{Insurance data}\label{ss:insurance-data}

We use a home insurance dataset from a large insurance company operating in Canada. We round the summary statistics such that they do not reveal customer information. The dataset contains information about individual perils, including \texttt{fire}, \texttt{other}, \texttt{water}, \texttt{wind}, \texttt{hail}, sewer backup (\texttt{SBU}) and \texttt{theft}. We do not know what type of coverages are included in the \texttt{other} peril. Finally, we create a category corresponding to the \texttt{total} perils. For each observation and each peril, we have the total frequency and total loss amounts. We consider the portion of the insurance portfolio located in Québec City, Québec, Canada, since this corresponds to the area where we have SVI. We have about 420,000 observations from 2009 to 2020, where the exposures for single observations vary from one day to one year (the mean and median exposures are respectively 0.44 and 0.42 years, meaning that most exposure periods are computed on a half-year basis). During this period, we possess SVI embeddings for around 40,000 distinct households, resulting in an approximate average of 10.5 observations per household.


Due to intellectual property constraints, we can only share relative data regarding the frequency and severity of various perils in our dataset. For each peril, the percentage of zeros is over 99\%, indicating that we are dealing with infrequent events. In terms of frequency, \texttt{water} perils are the most common, occurring approximately two to three times more frequently than \texttt{theft}, \texttt{other}, and \texttt{SBU} perils. The \texttt{wind} and \texttt{fire} perils are less common, with their frequency being 6 to 20 times lower than that of \texttt{water} perils. The least frequent peril is \texttt{hail}, which occurs 100 times less frequently than \texttt{water}. To analyze severity, we combine the \texttt{wind} and \texttt{hail} perils since there were too few observations to build a separate severity model for each. The median \texttt{total} severity is about 40\% smaller than the average \texttt{total} severity, indicating positive skewness. The \texttt{theft}, \texttt{other} and \texttt{wind \& hail} perils exhibit similar median severities. However, the median severities for \texttt{water}, \texttt{SBU} and \texttt{fire} perils are more than twice as high. The severity of fire risks is much more skewed towards higher values: even though the median severity for \texttt{fire} is similar to that of \texttt{water} and \texttt{SBU}, the average severity of \texttt{fire} is six times higher. This indicates that fire incidents may have many small damages but also very severe, for instance, in the case of the total destruction of the home. To limit the impact of outliers in the forthcoming gamma regression, we cap the severity at 100,000\$ in our severity models. We split the dataset into two parts: one for training the ratemaking models and the other for evaluating the model on out-of-sample observations. The training/testing split is done on a 90\% / 10\% basis. 


To demonstrate the seamless integration of our ratemaking framework with existing actuarial pricing systems, we also include conventional actuarial variables in our ratemaking models. In our case, we use four variables: the client's age (\texttt{cage}), the number of years since the roofing was updated (\texttt{roof}), the age of the building (\texttt{bage}) and the limit amount of building coverage (\texttt{limit}). In the dataset provided, the client's age is in fifteen buckets of five-year intervals; thus use dummy coding to construct features in the models. Building age was also a fine-tuning task (although this regression task had a high RMSE). 

\subsection{Frequency model}

We now perform the extrinsic evaluation of the embedding models, that is, evaluate if the image embeddings accomplish the task of capturing useful information for insurance pricing. We start with a GLM for frequency modelling, using a Poisson response with the canonical link function
\begin{equation}\label{eq:glm-poisson}
	\ln(E[Y]) = \beta_0 + \ln \omega + 
	\underbrace{\sum_{j = 1}^{p}x_{j} \alpha_{j}}_{\text{traditional component}} + 
	\underbrace{\sum_{k = 1}^{\ell}\gamma_{k} \beta_{k},}_{\text{embedding component}}
\end{equation}
where $\omega$ is the exposure based on annual exposure base, $p$ is the number of traditional features, $\ell$ is the embedding size for the image model, $x_j, j \in \{1,\dots, p\}$ are the traditional features and $\gamma_{k}$ is the $k$th embedding dimension, $k \in \{1, \dots, \ell\}$. 

We compare each image model architecture (DenseNet121 and ResNet 18, 50 and 101) along with a baseline model with no embedding component. For each image model, we compare using an embedding size of 8, 16 and 32. In Tables \ref{tab:results-FT}, \ref{tab:results-frozen} and \ref{tab:results-pca}, we present the deviance on the test set for the PCA, frozen and fine-tuned method of constructing embeddings. Recall that a smaller deviance means a higher likelihood, meaning the model fit is better. In parenthesis, we include the number of GLM parameters from the embedding component significantly different than zero at a 0.05 significance level. If embeddings learned feature spaces that were linearly independent of the canonical function of the response variable, one would expect the number of significant GLM parameters from the embedding component at a 0.05 significance level to be 0.4, 0.8 and 1.6 for embedding sizes 8, 16 and 32. We highlight in bold the number of significant GLM parameters at a 0.05 significance level above the expected number of significant parameters under the assumption of independence. In every case, adding embeddings decreases the training deviance, which is unsurprising since there are more degrees of freedom in the models with embeddings. To limit the length of this paper, we do not show the training deviance. 

\begin{table}[ht]
	\centering
	\resizebox{\textwidth}{!}{
		\begin{tabular}{llllllllll}
			Model                        & $\ell$ & \texttt{Theft}       & \texttt{Other}       & \texttt{SBU}         & \texttt{Water}       & \texttt{Hail}       & \texttt{Wind}       & \texttt{Fire}       & \texttt{Total}        \\ \hline
			Baseline                     & 0      & 1468.91              & 2028.72              & 2697.17              & 3831.07              & 163.47              & 884.08              & 479.96              & 8661.44               \\ \hline
			\multirow{3}{*}{ResNet18}    & 8      & 1468.06 (\textbf{1}) & 2029.27 (0)          & 2670.60 (\textbf{3}) & 3825.91 (\textbf{1}) & 166.78 (0)          & 881.25 (\textbf{1}) & 481.42 (\textbf{3}) & 8665.47 (\textbf{2})  \\
			& 16     & 1468.12 (\textbf{8}) & 2028.05 (\textbf{3}) & 2667.97 (\textbf{7}) & 3817.73 (\textbf{2}) & 165.56 (0)          & 884.93 (0)          & 477.01 (0)          & 8657.23 (\textbf{7})  \\
			& 32     & 1460.68 (\textbf{2}) & 2035.17 (\textbf{3}) & 2662.17 (\textbf{9}) & 3816.33 (\textbf{5}) & 182.68 (\textbf{4}) & 880.27 (0)          & 477.92 (\textbf{2}) & 8656.09 (\textbf{2})  \\ \hline
			\multirow{3}{*}{ResNet50}    & 8      & 1459.36 (\textbf{3}) & 2030.22 (0)          & 2692.66 (\textbf{3}) & 3825.50 (\textbf{1}) & 163.87 (0)          & 885.85 (\textbf{1}) & 477.11 (\textbf{3}) & 8670.48 (\textbf{3})  \\
			& 16     & 1463.73 (\textbf{2}) & 2028.17 (\textbf{2}) & 2661.89 (\textbf{9}) & 3820.33 (\textbf{7}) & 170.42 (0)          & 884.61 (0)          & 475.02 (0)          & 8658.43 (\textbf{4})  \\
			& 32     & 1462.73 (\textbf{7}) & 2029.40 (\textbf{3}) & 2661.36 (\textbf{8}) & 3816.36 (\textbf{2}) & 177.95 (0)          & 878.15 (\textbf{2}) & 481.98 (1)          & 8657.40 (\textbf{2})  \\ \hline
			\multirow{3}{*}{ResNet101}   & 8      & 1460.39 (\textbf{4}) & 2026.99 (\textbf{5}) & 2671.30 (\textbf{5}) & 3821.99 (0)          & 168.80 (0)          & 878.75 (\textbf{3}) & 477.50 (\textbf{1}) & 8659.38 (\textbf{3})  \\
			& 16     & 1464.28 (\textbf{7}) & 2028.09 (0)          & 2663.49 (\textbf{6}) & 3820.84 (\textbf{3}) & 169.21 (0)          & 883.01 (\textbf{1}) & 474.28 (0)          & 8659.49 (\textbf{6})  \\
			& 32     & 1468.98 (\textbf{3}) & 2030.89 (\textbf{2}) & 2665.01 (\textbf{8}) & 3819.22 (\textbf{4}) & 173.32 (\textbf{2}) & 893.89 (1)          & 478.95 (1)          & 8664.47 (\textbf{10}) \\ \hline
			\multirow{3}{*}{DenseNet121} & 8      & 1473.56 (\textbf{5}) & 2028.77 (\textbf{2}) & 2672.94 (\textbf{5}) & 3823.37 (\textbf{1}) & 166.79 (\textbf{1}) & 879.85 (\textbf{1}) & 477.91 (0)          & 8664.57 (\textbf{2})  \\
			& 16     & 1467.33 (\textbf{2}) & 2031.46 (\textbf{4}) & 2665.68 (\textbf{6}) & 3826.43 (0)          & 174.70 (0)          & 878.79 (\textbf{1}) & 477.26 (0)          & 8661.09 (\textbf{4})  \\
			& 32     & 1477.16 (\textbf{6}) & 2024.60 (\textbf{3}) & 2668.62 (\textbf{7}) & 3823.74 (\textbf{7}) & 173.28 (0)          & 885.86 (0)          & 479.82 (0)          & 8660.95 (\textbf{5})  \\ \hline
		\end{tabular}					
	}
	\caption{Testing deviance for frequency prediction with fine-tuned models.}\label{tab:results-FT}
\end{table}
\begin{table}[ht]
	\centering
	\resizebox{\textwidth}{!}{
	\begin{tabular}{llllllllll}
		Model                        & $\ell$ & \texttt{Theft}       & \texttt{Other}       & \texttt{SBU}         & \texttt{Water}       & \texttt{Hail}       & \texttt{Wind}       & \texttt{Fire}       & \texttt{Total}       \\ \hline
		Baseline                     & 0      & 1468.91              & 2028.72              & 2697.17              & 3831.07              & 163.47              & 884.08              & 479.96              & 8661.44              \\ \hline
		\multirow{3}{*}{ResNet18}    & 8      & 1469.18 (\textbf{1}) & 2031.90 (0)          & 2683.62 (\textbf{2}) & 3828.06 (0)          & 161.78 (\textbf{1}) & 882.82 (0)          & 472.73 (\textbf{1}) & 8662.96 (\textbf{1}) \\
		                             & 16     & 1472.32 (\textbf{1}) & 2027.60 (0)          & 2682.79 (\textbf{1}) & 3824.34 (\textbf{4}) & 162.67 (\textbf{1}) & 885.53 (\textbf{1}) & 475.82 (0)          & 8660.15 (\textbf{4}) \\
		                             & 32     & 1476.20 (1)          & 2032.81 (0)          & 2674.39 (\textbf{5}) & 3830.33 (\textbf{3}) & 175.35 (\textbf{2}) & 883.74 (\textbf{2}) & 466.54 (\textbf{3}) & 8660.23 (\textbf{4}) \\ \hline
		\multirow{3}{*}{ResNet50}    & 8      & 1469.59 (\textbf{1}) & 2031.52 (0)          & 2681.21 (\textbf{4}) & 3824.65 (\textbf{1}) & 165.43 (0)          & 885.21 (0)          & 474.84 (0)          & 8661.91 (\textbf{2}) \\
		                             & 16     & 1472.47 (\textbf{3}) & 2030.93 (\textbf{1}) & 2677.99 (\textbf{2}) & 3824.71 (\textbf{1}) & 171.37 (0)          & 886.23 (\textbf{2}) & 478.07 (\textbf{2}) & 8659.54 (\textbf{1}) \\
		                             & 32     & 1477.54 (1)          & 2038.33 (\textbf{3}) & 2679.71 (0)          & 3829.98 (1)          & 176.22 (\textbf{2}) & 886.11 (\textbf{2}) & 475.72 (0)          & 8665.00 (0)          \\ \hline
		\multirow{3}{*}{ResNet101}   & 8      & 1471.24 (\textbf{2}) & 2031.64 (0)          & 2686.02 (\textbf{1}) & 3823.83 (0)          & 169.96 (\textbf{3}) & 883.29 (0)          & 477.17 (0)          & 8665.15 (0)          \\
		                             & 16     & 1472.27 (\textbf{4}) & 2031.03 (\textbf{2}) & 2682.60 (\textbf{1}) & 3823.30 (\textbf{1}) & 169.02 (0)          & 881.50 (0)          & 475.97 (\textbf{1}) & 8661.69 (\textbf{3}) \\
		                             & 32     & 1472.38 (0)          & 2028.80 (0)          & 2677.00 (\textbf{3}) & 3828.52 (\textbf{5}) & 177.78 (\textbf{3}) & 890.98 (\textbf{2}) & 480.66 (\textbf{4}) & 8658.90 (1)          \\ \hline
		\multirow{3}{*}{DenseNet121} & 8      & 1466.73 (\textbf{1}) & 2030.89 (0)          & 2686.41 (\textbf{2}) & 3826.29 (0)          & 164.15 (0)          & 883.90 (0)          & 477.74 (\textbf{1}) & 8663.47 (\textbf{1}) \\
		                             & 16     & 1467.48 (\textbf{1}) & 2034.43 (0)          & 2680.16 (\textbf{3}) & 3825.57 (\textbf{3}) & 160.94 (\textbf{2}) & 887.59 (0)          & 480.37 (0)          & 8660.13 (\textbf{2}) \\
		                             & 32     & 1466.29 (\textbf{2}) & 2038.45 (1)          & 2678.05 (\textbf{4}) & 3829.30 (1)          & 164.56 (\textbf{2}) & 889.04 (0)          & 484.92 (1)          & 8664.46 (\textbf{5}) \\ \hline
	\end{tabular}}
	\caption{Testing deviance for frequency prediction with frozen models.}\label{tab:results-frozen}
\end{table}
\begin{table}[ht]
	\centering
	\resizebox{\textwidth}{!}{
		\begin{tabular}{llllllllll}
			Model                        & $\ell$ & \texttt{Theft}       & \texttt{Other}       & \texttt{SBU}          & \texttt{Water}       & \texttt{Hail}       & \texttt{Wind}       & \texttt{Fire}       & \texttt{Total}       \\ \hline
			Baseline                     & 0      & 1468.91              & 2028.72              & 2697.17               & 3831.07              & 163.47              & 884.08              & 479.96              & 8661.44              \\ \hline
			\multirow{3}{*}{ResNet18}    & 8      & 1467.13 (0)          & 2029.41 (\textbf{1}) & 2687.35 (\textbf{6})  & 3824.90 (\textbf{2}) & 162.36 (\textbf{1}) & 882.65 (\textbf{2}) & 470.51 (\textbf{1}) & 8659.98 (\textbf{3}) \\
			& 16     & 1470.41 (0)          & 2028.73 (\textbf{1}) & 2680.71 (\textbf{9})  & 3823.31 (\textbf{2}) & 163.68 (\textbf{2}) & 881.86 (\textbf{3}) & 468.21 (\textbf{1}) & 8658.23 (\textbf{3}) \\
			& 32     & 1477.80 (\textbf{5}) & 2030.89 (\textbf{3}) & 2680.65 (\textbf{12}) & 3827.10 (\textbf{2}) & 172.65 (\textbf{4}) & 879.11 (\textbf{4}) & 466.83 (\textbf{2}) & 8663.76 (\textbf{4}) \\ \hline
			\multirow{3}{*}{ResNet50}    & 8      & 1465.87 (\textbf{1}) & 2029.84 (0)          & 2684.10 (\textbf{5})  & 3827.18 (0)          & 169.44 (\textbf{2}) & 882.71 (\textbf{2}) & 472.90 (\textbf{1}) & 8659.97 (\textbf{2}) \\
			& 16     & 1467.17 (\textbf{2}) & 2029.90 (0)          & 2678.30 (\textbf{8})  & 3830.13 (\textbf{2}) & 171.98 (\textbf{3}) & 881.84 (\textbf{3}) & 471.76 (\textbf{1}) & 8663.14 (\textbf{4}) \\
			& 32     & 1471.53 (\textbf{4}) & 2037.13 (\textbf{2}) & 2677.54 (\textbf{12}) & 3826.82 (\textbf{2}) & 177.48 (\textbf{5}) & 883.00 (\textbf{4}) & 473.82 (\textbf{3}) & 8662.24 (\textbf{6}) \\ \hline
			\multirow{3}{*}{ResNet101}   & 8      & 1466.68 (0)          & 2028.41 (\textbf{1}) & 2691.01 (\textbf{3})  & 3828.88 (\textbf{2}) & 163.32 (\textbf{1}) & 883.10 (\textbf{1}) & 472.99 (\textbf{1}) & 8662.80 (0)          \\
			& 16     & 1473.02 (\textbf{3}) & 2029.71 (\textbf{1}) & 2686.43 (\textbf{8})  & 3828.70 (\textbf{3}) & 171.36 (\textbf{3}) & 884.29 (\textbf{2}) & 475.40 (\textbf{1}) & 8661.85 (\textbf{1}) \\
			& 32     & 1478.86 (\textbf{5}) & 2033.07 (1)          & 2678.64 (\textbf{13}) & 3824.33 (\textbf{5}) & 168.79 (\textbf{4}) & 880.54 (\textbf{3}) & 473.26 (1)          & 8654.87 (\textbf{4}) \\ \hline
			\multirow{3}{*}{DenseNet121} & 8      & 1468.00 (\textbf{1}) & 2031.26 (0)          & 2693.71 (\textbf{5})  & 3827.15 (\textbf{1}) & 164.41 (\textbf{1}) & 883.80 (\textbf{1}) & 479.15 (\textbf{2}) & 8665.43 (\textbf{2}) \\
			& 16     & 1467.77 (\textbf{1}) & 2037.79 (0)          & 2688.06 (\textbf{6})  & 3824.19 (\textbf{1}) & 162.47 (\textbf{1}) & 884.06 (\textbf{1}) & 477.76 (\textbf{1}) & 8657.18 (\textbf{2}) \\
			& 32     & 1474.54 (\textbf{5}) & 2041.15 (1)          & 2687.36 (\textbf{9})  & 3826.71 (\textbf{3}) & 165.85 (\textbf{1}) & 881.89 (1)          & 476.55 (\textbf{2}) & 8658.33 (\textbf{5}) \\ \hline
	\end{tabular}}
	\caption{Testing deviance for frequency prediction with principal components.}\label{tab:results-pca}
\end{table}

What is immediately noticeable from Tables \ref{tab:results-FT}, \ref{tab:results-frozen} and \ref{tab:results-pca} is that the embeddings greatly reduce the deviance for \texttt{SBU} and \texttt{water} claim frequency models. Increasing the embedding size provides better performance, meaning that the higher embedding dimensions do not learn redundant information (this point is even more valid for the frozen and fine-tuned models). For \texttt{SBU} and \texttt{water} perils, increasing the flexibility of the representation model leads to a better quality of representations to model claim frequency: for a fixed embedding size and model, going from PCA to frozen to fine-tuned models generally leads to better results, meaning that adapting the weights of the image models to focus on characteristics of the house leads to better representations of \texttt{SBU} and \texttt{water} damage claim frequency. 

For the remaining perils, the results are more disputable. In most cases, we observe a slight increase or decrease in the deviance, so it isn't clear if the embeddings capture a true effect within the SVI or if this is due to randomness. For instance, looking at the \texttt{theft} peril in the PCA approach, increasing the embedding dimensions generally leads to an increase in deviance, meaning that the GLM may have overfit on the higher embedding dimensions. The best model for the \texttt{fire} peril is the PCA with the ResNet18 model. The fine-tuned models to predict the frequency of \texttt{fire} perils do not appear to increase or decrease the deviance systematically. Therefore, the related tasks we selected from the property tax data may not be related to the occurrence of fires. Most ResNet models from the fine-tuned approach for the \texttt{theft} and \texttt{wind} peril lead to decreased deviance compared to the baseline model, but there is no clear-cut relationship between the embedding dimension and the deviance. This ambiguity makes understanding the role of the embeddings in the model harder, so someone might choose not to use them.

For every model, the \texttt{other} and \texttt{total} perils, there is no single model that is either better or worse than the baseline, implying that the embeddings are not useful within the GLM model. We find this surprising: the embeddings unquestionably improved the \texttt{SBU} and \texttt{water} perils, which are the most common ones in the dataset. However, it seems that combining the remaining perils into one regression task eliminates that improvement. We, therefore, stress that embeddings have different effects on each peril because the embeddings are not useful for the ``catch-all'' perils \texttt{other} and \texttt{total}. 

The only peril for which embeddings usually increase the deviance is \texttt{hail}. This peril occurs when a hailstorm causes damage to a house, say, if a large hailstone falls on a roof or shatters a window. The increase in deviance may be due to a lack of causal effect (there may be no link between the picture of a house and the probability of filing a claim due to hail). Another explanation is that the hail peril is the rarest in the dataset: hailstorms are infrequent in Québec city (see, for instance, \cite{etkin2018hail} for some frequency statistics), and when they do occur, the size of hailstones is typically too small to cause a loss larger than the deductible amount. 

When looking at the number of significant parameters at the 0.05 level for the embedding component, one observes that this number, for most models, is higher than the expected number under the independence assumption. For instance, for the fine-tuned models, almost all of the models for \texttt{theft}, \texttt{SBU}, \texttt{water}, and \texttt{total} perils had a high number of significant GLM parameters for the embedding component, even if we could only conclude that the \texttt{SBU} and the \texttt{water} claims had clear improvements in the models. Further, for the \texttt{SBU} peril, one usually has more significant parameters for the PCA approach compared with the fine-tuned approach, even though the fine-tuned yields smaller deviance values. Therefore, the statistical significance of the embedding parameters may not be a good extrinsic evaluation measure for embeddings. However, it would be cause for alarm if a model with embeddings yielded a much smaller deviance value if none of the parameters associated with the embedding components were statistically significant. 

\subsection{Effect of including variables in both relevant tasks and traditional variables}

Let us next look at the perils for which the SVI embeddings provided the best improvements and their effect on the age of the building. Recall that the age of the building was a fine-tuning task from the tax assessments; hence the image representations with the frozen and fine-tuned approaches capture representations of the age (even if the RMSE of these models was high). Therefore, the model containing embeddings constructed with the frozen and fine-tuned approaches uses both the \texttt{bage} variable and a proxy to the \texttt{bage} variable through the unsupervised transfer learning. Note, however, that the variable in the insurance dataset is dynamic (age changes every year since we have observations between 2009 and 2020), while the proxy in the embeddings is static (since it depends on the age of the building at the tax assessment date). In Table \ref{tab:p-value-age-building}, we present the $p$-value associated with the variable \texttt{bage} for the baseline model and the three embedding construction approaches. Note that the \texttt{bage} variable is significant in both models (assuming a 0.05 significance level). We consider only ResNet18 and ResNet101 models with an embedding dimension of 32 to limit space, but one obtains similar results for other embedding sizes and image model architectures. Further, we only consider the \texttt{SBU} and \texttt{water} perils since they are the ones for which SVI embeddings improved the deviance for every set of embedding. For the \texttt{SBU} peril, including the embeddings, does not deem the \texttt{bage} variable insignificant since they remain under the 0.05 threshold for every embedding construction method, image model architecture and embedding size. For the \texttt{water} peril, including any embedding makes the \texttt{bage} variable insignificant, meaning that the embeddings may be a better proxy to any risk-generating process than the age of the building. We remark that the $p$-values for the frozen and fine-tuned model are higher than for the PCA models, but one should not confuse this with the notion of having ``less significance'' but that no effect was observed. From this experiment, we cannot conclude if the image model or the unsupervised transfer learning component of the representation learning framework yielded the \texttt{bage} insignificant for the \texttt{water} peril. 
\begin{table}[ht]
	\centering
	\begin{tabular}{cccccccc}
		               &  Baseline   &  \multicolumn{2}{c}{PCA}  & \multicolumn{2}{c}{frozen} & \multicolumn{2}{c}{fine-tuned} \\
		               		\cmidrule(r){3-4}\cmidrule(r){5-6}\cmidrule(r){7-8}
		               &             &  ResNet18   &  ResNet32   &  ResNet18   &   ResNet32   & ResNet18  &      ResNet32      \\
		 \texttt{SBU}  & $<10^{-10}$ & $<10^{-10}$ & $<10^{-10}$ & $<10^{-10}$ & $<10^{-10}$  & 0.0005912 &      0.000018      \\
		\texttt{water} &  0.006181   &  0.312924   &  0.280731   &  0.390484   &   0.943965   & 0.445789  &      0.704342\\\hline
	\end{tabular}
	\caption{Comparison of $p$-values for the variable \texttt{bage} with and without embeddings.}\label{tab:p-value-age-building}
\end{table}

Another diagnostic that will help us interpret the results of the regression is the variance inflation factors (VIF), more specifically, their generalized version introduced in \cite{fox1992generalized} since the \texttt{cage} attribute contains 14 degrees of freedom. In Table \ref{tab:VIF}, we present the VIF for the baseline model and different embedding construction methods for the ResNet18 model with eight embedding dimensions. We also present results for models trained with the embedding component only, that is, without the traditional component. 

\begin{table}[ht]
	\centering
	\resizebox{\textwidth}{!}{
		\begin{tabular}{lrrrrrrrrrrrrr}
			                     & offset & \texttt{cage} & \texttt{roof} & \texttt{bage} & \texttt{limit} & $\gamma_1$ & $\gamma_2$ & $\gamma_3$ & $\gamma_4$ & $\gamma_5$ & $\gamma_6$ & $\gamma_7$ & $\gamma_8$ \\ \hline
			Baseline             &  1.006 &         1.080 &         1.018 &         1.116 &          1.043 &            &            &            &            &            &            &            &            \\ \hline
			PCA                  &  1.000 &               &               &               &                &      1.274 &      1.172 &      1.031 &      1.019 &      1.034 &      1.155 &      1.003 &      1.043 \\
			                     &  1.007 &         1.098 &         1.018 &         1.273 &          1.097 &      1.302 &      1.263 &      1.045 &      1.057 &      1.055 &      1.158 &      1.008 &      1.050 \\ \hline
			Frozen               &  1.000 &               &               &               &                &      7.215 &     38.938 &     14.329 &     13.733 &     26.833 &     54.160 &      5.109 &     16.064 \\
			                     &  1.007 &         1.101 &         1.017 &         1.610 &          1.153 &      7.323 &     40.793 &     14.467 &     13.846 &     27.622 &     55.760 &      5.996 &     16.296 \\ \hline
			Fine-tuned           &  1.000 &               &               &               &                &     57.250 &     23.677 &     60.947 &      6.662 &      7.675 &    104.657 &     55.757 &     34.132 \\
			                     &  1.011 &         1.120 &         1.017 &         4.410 &          2.084 &     58.914 &     31.823 &     63.224 &      6.887 &      8.100 &    106.449 &     62.353 &     52.510 \\ \hline
		\end{tabular}}
	\caption{Variance inflation factors for different embedding construction approaches.}\label{tab:VIF}
\end{table}

Recall that a VIF over ten is considered problematic and that a cutoff of 5 is often recommended. The VIFs in the baseline model are all around one, indicating low collinearity. The VIF for each variable in the traditional component usually increases when adding the embedding component and vice versa. Let us now use VIFs to examine the effect of including a variable in the regression model and the related tasks. The VIF for \texttt{bage} increases as the embedding model increases. For the most flexible embedding construction method, the VIF is about four times as large as the baseline, indicating moderate collinearity and hinting that the \texttt{bage} variable may become duplicated within the embedding dimensions. Note that the VIF for the embedding components is very high: we will examine the reasons and propose a solution in the following subsection. 

\subsection{Impact of correlated embeddings}

\begin{figure}[ht]
	\centering
	\begin{tikzpicture}[scale = 0.9]
		\begin{axis}[
			axis equal image,
			scatter,
			colormap={rwb}{color=(red) color=(white) color=(blue)},
			colorbar,
			point meta min=-1,
			point meta max=1,
			minor tick num=1,
			tickwidth=0pt,
			y dir=reverse,
			xticklabels={},
			yticklabels={},
			extra x ticks={0,1, ..., 7},
			extra x tick labels={$\gamma_1$, $\gamma_2$, $\gamma_3$, $\gamma_4$, $\gamma_5$, $\gamma_6$, $\gamma_7$, $\gamma_8$},
			extra y ticks={0, 1, ..., 7},
			extra y tick labels={$\gamma_1$, $\gamma_2$, $\gamma_3$, $\gamma_4$, $\gamma_5$, $\gamma_6$, $\gamma_7$, $\gamma_8$},
			xticklabel pos=right,
			enlargelimits={abs=0.5},
			scatter/@pre marker code/.append code={
				\pgfplotstransformcoordinatex{sqrt(abs(\pgfplotspointmeta))}
				\scope[mark size=\pgfplotsunitxlength*\pgfmathresult/2, fill=mapped color]
			},
			scatter/@post marker code/.append code={
				\endscope
			}
			]
			\addplot +[
			point meta=explicit,
			only marks,
			] table [
			x expr={int(mod(\coordindex+0.01,8))},
			y expr={int((\coordindex+0.01)/8))},
			meta=value
			] {
				X   Y   value
				1   1   1
				2   1   0.1153
				3   1   0.0812
				4   1   -0.1386
				5   1   0.0803
				6   1   0.0735
				7   1   0.1396
				8   1   0.0195
				1   2   0.1153
				2   2   1
				3   2   0.2495
				4   2   -0.1308
				5   2   -0.2245
				6   2   0.3325
				7   2   0.1087
				8   2   0.285
				1   3   0.0812
				2   3   0.2495
				3   3   1
				4   3   -0.4432
				5   3   0.6464
				6   3   0.9237
				7   3   0.826
				8   3   0.8262
				1   4   -0.1386
				2   4   -0.1308
				3   4   -0.4432
				4   4   1
				5   4   -0.5841
				6   4   -0.1912
				7   4   -0.8163
				8   4   -0.3782
				1   5   0.0803
				2   5   -0.2245
				3   5   0.6464
				4   5   -0.5841
				5   5   1
				6   5   0.4354
				7   5   0.8002
				8   5   0.424
				1   6   0.0735
				2   6   0.3325
				3   6   0.9237
				4   6   -0.1912
				5   6   0.4354
				6   6   1
				7   6   0.6598
				8   6   0.6966
				1   7   0.1396
				2   7   0.1087
				3   7   0.826
				4   7   -0.8163
				5   7   0.8002
				6   7   0.6598
				7   7   1
				8   7   0.5811
				1   8   0.0195
				2   8   0.285
				3   8   0.8262
				4   8   -0.3782
				5   8   0.424
				6   8   0.6966
				7   8   0.5811
				8   8   1
			};
		\end{axis}
	\end{tikzpicture}
	\quad 
	\begin{tikzpicture}[scale = 0.9]
		\begin{axis}[
			axis equal image,
			scatter,
			colormap={rwb}{color=(red) color=(white)
				color=(blue)},
			colorbar,
			point meta min=-1,
			point meta max=1,
			minor tick num=1,
			tickwidth=0pt,
			y dir=reverse,
			xticklabels={},
			yticklabels={},
			extra x ticks={0,1, ..., 7},
			extra x tick labels={$\gamma_1$, $\gamma_2$, $\gamma_3$, $\gamma_4$, $\gamma_5$, $\gamma_6$, $\gamma_7$, $\gamma_8$},
			extra y ticks={0, 1, ..., 7},
			extra y tick labels={$\gamma_1$, $\gamma_2$, $\gamma_3$, $\gamma_4$, $\gamma_5$, $\gamma_6$, $\gamma_7$, $\gamma_8$},
			xticklabel pos=right,
			enlargelimits={abs=0.5},
			scatter/@pre marker code/.append code={
				\pgfplotstransformcoordinatex{sqrt(abs(\pgfplotspointmeta))}
				\scope[mark size=\pgfplotsunitxlength*\pgfmathresult/2, fill=mapped color]
			},
			scatter/@post marker code/.append code={
				\endscope
			}
			]
			\addplot +[
			point meta=explicit,
			only marks,
			] table [
			x expr={int(mod(\coordindex+0.01,8))},
			y expr={int((\coordindex+0.01)/8))},
			meta=value
			] {
				X   Y   value
				
				1   1   1
				2   1   -0.0826
				3   1   0.1318
				4   1   -0.0943
				5   1   0.017
				6   1   -0.1043
				7   1   -0.0968
				8   1   -0.1044
				1   2   -0.0826
				2   2   1
				3   2   -0.8828
				4   2   0.5625
				5   2   -0.4435
				6   2   0.1053
				7   2   0.4145
				8   2   0.8933
				1   3   0.1318
				2   3   -0.8828
				3   3   1
				4   3   -0.5356
				5   3   0.465
				6   3   -0.2894
				7   3   -0.3757
				8   3   -0.8879
				1   4   -0.0943
				2   4   0.5625
				3   4   -0.5356
				4   4   1
				5   4   -0.2734
				6   4   0.2269
				7   4   0.9622
				8   4   0.657
				1   5   0.017
				2   5   -0.4435
				3   5   0.465
				4   5   -0.2734
				5   5   1
				6   5   -0.7771
				7   5   -0.1967
				8   5   -0.6283
				1   6   -0.1043
				2   6   0.1053
				3   6   -0.2894
				4   6   0.2269
				5   6   -0.7771
				6   6   1
				7   6   0.1824
				8   6   0.3387
				1   7   -0.0968
				2   7   0.4145
				3   7   -0.3757
				4   7   0.9622
				5   7   -0.1967
				6   7   0.1824
				7   7   1
				8   7   0.5581
				1   8   -0.1044
				2   8   0.8933
				3   8   -0.8879
				4   8   0.657
				5   8   -0.6283
				6   8   0.3387
				7   8   0.5581
				8   8   1
				
			};
		\end{axis}
	\end{tikzpicture}

	\caption{Correlation matrix of embedding dimensions for ResNet-18 with 8 embeddings for frozen (left) and fine-tuned (right) approaches.}\label{fig:corr-matrix-emb}
\end{figure}
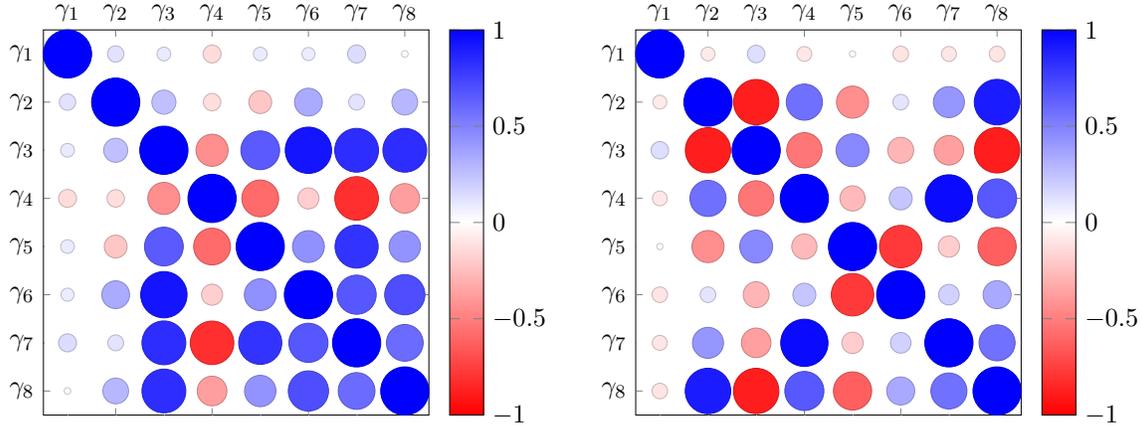

One aspect to consider for applications of embeddings is the correlation of embedding dimensions. If the correlation is too high, there may be collinearity in the features, which could cause the variance of regression coefficients to be inflated (as observed in Table \ref{tab:VIF}). Note that, by construction, PCA embeddings are orthogonal to each other (hence linearly uncorrelated), so we must only diagnose the correlation of the embeddings generated by unsupervised transfer learning. In Figure \ref{fig:corr-matrix-emb}, we present the correlation matrix for the ResNet18 embeddings with eight embedding dimensions for the frozen and fine-tuned approaches of embedding construction. One observes a linear correlation (between -0.75 and 0.91) between the embedding dimensions. One way to remove this linear correlation is to use all of the principal components of the embeddings instead of the embeddings themselves. While this approach does not make much sense in typical GLM modelling (since one would lose the ability to interpret the model and perform variable selection), doing so on image embeddings does not hurt the model since the image embeddings are already uninterpretable. 

In Table \ref{tab:VIF2}, we present the VIFs for the GLM trained on decorrelated embedding dimensions. One observes that the VIF for the traditional component stays the same if one uses the frozen/fined-tuned embeddings or their principal components. The VIFs in Table \ref{tab:VIF} for the frozen and fine-tuned embedding models are high, always over the cutoff of 5 and sometimes reaching over 100. However, when using the principal components of the frozen and fine-tuned embeddings, the VIFs are no longer considered high. It follows that the variance of the predictors is lower, implying that more variables become statistically significant. Note that this step does not impact the predictive variance but provides a more useful way to diagnose the statistical significance of embedding parameters and the effect of the collinearity of the embeddings on the GLM. 

\begin{table}[ht]
	\centering
	\resizebox{\textwidth}{!}{
		\begin{tabular}{lrrrrrrrrrrrrr}
			                 & offset & \texttt{cage} & \texttt{roof} & \texttt{bage} & \texttt{limit} & $\gamma_1$ & $\gamma_2$ & $\gamma_3$ & $\gamma_4$ & $\gamma_5$ & $\gamma_6$ & $\gamma_7$ & $\gamma_8$ \\ \hline
			Frozen + PCA     &  1.000 &               &               &               &                &      1.848 &      2.186 &      1.080 &      1.345 &      2.400 &      1.835 &      1.185 &      1.320 \\
			                 &  1.007 &         1.101 &         1.017 &         1.610 &          1.153 &      1.912 &      2.344 &      1.129 &      1.708 &      2.441 &      1.828 &      1.186 &      1.312 \\ \hline
			Fine-tuned + PCA &  1.000 &               &               &               &                &      1.717 &      1.963 &      2.493 &      2.276 &      1.280 &      1.448 &      3.096 &      1.302 \\
			                 &  1.011 &         1.120 &         1.017 &         4.410 &          2.084 &      1.943 &      2.109 &      5.053 &      3.094 &      2.448 &      1.535 &      3.219 &      1.313\\\hline
		\end{tabular}}
	\caption{Variance inflation factors after decorrelating embeddings.}\label{tab:VIF2}
\end{table}

\subsection{Severity model}

We next extrinsically evaluate the quality of severity models. In this case, we use a gamma response with the canonical link function: 
\begin{equation}\label{eq:glm-gamma}
	E[Y]^{-1} = \beta_0 + \sum_{j = 1}^{p}x_{j} \alpha_{j} + 
	\sum_{k = 1}^{\ell}\gamma_{k} \beta_{k}.
\end{equation}
In Tables 
\ref{tab:results-sev-FT}, 
\ref{tab:results-sev-frozen} and
\ref{tab:results-sev-pca},
we present deviance results on the test dataset with the PCA, frozen and fine-tuned approaches respectively. Recall that since the number of \texttt{wind} and \texttt{hail} perils was too small, we trained a combined model for these two perils. Further, some GLM models did not converge for the \texttt{fire} peril with the PCA approach to construct embeddings; we denote these by NA in the deviance result tables. 

\begin{table}[ht]
	\centering
	\resizebox{\textwidth}{!}{
		\begin{tabular}{llrrrrrrr}
			Model                        & $\ell$ &      \texttt{Theft} &      \texttt{Other} &        \texttt{SBU} &      \texttt{Water} & \texttt{Wind} \& \texttt{Hail} &       \texttt{Fire} &       \texttt{Total} \\ \hline
			Baseline                     & 0      &               95.24 &              225.71 &              298.61 &              521.48 &                          46.74 &               47.28 &              1220.89 \\ \hline
			\multirow{3}{*}{ResNet18}    & 8      &          106.03 (0) &          231.51 (0) &          296.43 (0) &          520.11 (0) &                      46.97 (0) &  75.23 (\textbf{1}) &          1225.91 (0) \\
			& 16     & 107.42 (\textbf{2}) & 230.90 (\textbf{2}) & 297.50 (\textbf{2}) &          520.24 (0) &             49.71 (\textbf{1}) & 194.07 (\textbf{1}) &          1216.52 (0) \\
			& 32     & 105.78 (\textbf{2}) & 240.60 (\textbf{2}) & 296.68 (\textbf{2}) &          523.33 (0) &             60.16 (\textbf{5}) & 120.04 (\textbf{2}) & 1226.94 (\textbf{4}) \\ \hline
			\multirow{3}{*}{ResNet50}    & 8      &          107.29 (0) &          226.51 (0) &          293.31 (0) &          521.60 (0) &                      47.01 (0) &  56.13 (\textbf{2}) &          1223.54 (0) \\
			& 16     & 103.91 (\textbf{1}) & 223.79 (\textbf{3}) & 295.71 (\textbf{1}) & 517.79 (\textbf{1}) &                      50.13 (0) &  67.36 (\textbf{1}) & 1230.12 (\textbf{5}) \\
			& 32     &          104.82 (1) & 227.82 (\textbf{8}) &          294.36 (1) & 534.31 (\textbf{2}) &                      51.13 (0) &           64.48 (0) & 1224.14 (\textbf{2}) \\ \hline
			\multirow{3}{*}{ResNet101}   & 8      &           99.50 (0) &          225.43 (0) &          293.50 (0) &          519.70 (0) &                      48.59 (0) & 102.99 (\textbf{1}) &          1227.88 (0) \\
			& 16     &  98.92 (\textbf{1}) & 228.82 (\textbf{1}) & 291.73 (\textbf{1}) & 523.08 (\textbf{2}) &             48.75 (\textbf{2}) &  77.04 (\textbf{4}) & 1227.89 (\textbf{1}) \\
			& 32     &          108.00 (1) &          222.63 (0) & 293.31 (\textbf{6}) & 526.59 (\textbf{2}) &                      45.69 (1) & 307.10 (\textbf{5}) &          1240.69 (1) \\ \hline
			\multirow{3}{*}{DenseNet121} & 8      &  98.31 (\textbf{2}) & 230.58 (\textbf{1}) &          293.14 (0) &          518.43 (0) &                      47.85 (0) &           55.11 (0) & 1217.55 (\textbf{2}) \\
			& 16     & 102.45 (\textbf{1}) &          228.88 (0) & 291.90 (\textbf{2}) &          519.27 (0) &             46.95 (\textbf{1}) &  63.48 (\textbf{1}) & 1228.68 (\textbf{2}) \\
			& 32     &           92.31 (0) & 244.53 (\textbf{8}) & 293.50 (\textbf{3}) & 525.44 (\textbf{2}) &                      53.29 (0) & 112.33 (\textbf{2}) &          1228.05 (0) \\ \hline
		\end{tabular}					
	}
	\caption{Testing deviance for severity prediction with fine-tuned models.}\label{tab:results-sev-FT}
\end{table}

\begin{table}[ht]
	\centering
	\resizebox{\textwidth}{!}{
		\begin{tabular}{llrrrrrrr}
			Model                        & $\ell$ &      \texttt{Theft} &      \texttt{Other} &        \texttt{SBU} &      \texttt{Water} & \texttt{Wind} \& \texttt{Hail} &      \texttt{Fire} &       \texttt{Total} \\ \hline
			Baseline                     & 0      &               95.24 &              225.71 &              298.61 &              521.48 &                          46.74 &              47.28 &              1220.89 \\ \hline
			\multirow{3}{*}{ResNet18}    & 8      &           97.34 (0) & 221.27 (\textbf{1}) & 292.41 (\textbf{1}) &          521.37 (0) &                      46.10 (0) & 45.38 (\textbf{6}) & 1220.12 (\textbf{1}) \\
			& 16     & 102.85 (\textbf{1}) & 223.97 (\textbf{2}) & 288.03 (\textbf{4}) &          519.49 (0) &                      49.28 (0) & 40.67 (\textbf{1}) & 1216.13 (\textbf{1}) \\
			& 32     &  94.56 (\textbf{2}) &          225.00 (0) & 295.39 (\textbf{2}) &          519.08 (0) &                      60.51 (9) &          69.46 (1) &          1215.95 (0) \\ \hline
			\multirow{3}{*}{ResNet50}    & 8      &           92.96 (0) & 233.08 (\textbf{1}) & 297.06 (\textbf{1}) &          516.06 (0) &                      50.16 (0) &          53.33 (0) & 1221.52 (\textbf{2}) \\
			& 16     &           96.56 (0) &          229.65 (0) & 297.46 (\textbf{2}) &          516.49 (0) &                      52.40 (0) & 61.37 (\textbf{4}) & 1221.67 (\textbf{1}) \\
			& 32     & 105.37 (\textbf{3}) & 242.79 (\textbf{5}) & 296.65 (\textbf{2}) &          525.25 (1) &             54.33 (\textbf{3}) & 87.14 (\textbf{4}) &          1221.94 (0) \\ \hline
			\multirow{3}{*}{ResNet101}   & 8      &           95.73 (0) &          230.53 (0) &          298.55 (0) &          519.11 (0) &                      53.48 (0) &          50.11 (0) &          1218.42 (0) \\
			& 16     &           97.68 (0) & 224.30 (\textbf{1}) & 299.70 (\textbf{2}) &          514.68 (0) &             53.72 (\textbf{2}) &          42.55 (0) & 1222.20 (\textbf{1}) \\
			& 32     &  96.85 (\textbf{4}) & 238.12 (\textbf{6}) &          299.13 (0) & 520.12 (\textbf{2}) &             55.85 (\textbf{2}) & 43.36 (\textbf{2}) &          1214.32 (1) \\ \hline
			\multirow{3}{*}{DenseNet121} & 8      &           96.85 (0) & 232.33 (\textbf{2}) & 298.37 (\textbf{1}) &          518.93 (0) &                      49.26 (0) &          49.92 (0) &          1226.32 (0) \\
			& 16     &           97.15 (0) & 238.72 (\textbf{3}) & 297.81 (\textbf{1}) &          520.40 (0) &                      51.92 (0) & 64.97 (\textbf{3}) & 1220.43 (\textbf{1}) \\
			& 32     &          101.36 (1) & 234.20 (\textbf{3}) & 295.07 (\textbf{2}) & 510.46 (\textbf{2}) &                      61.74 (0) & 58.86 (\textbf{4}) & 1229.52 (\textbf{2}) \\ \hline
	\end{tabular}}
	\caption{Testing deviance for severity prediction with frozen models.}\label{tab:results-sev-frozen}
\end{table}

\begin{table}[ht]
	\centering
	\resizebox{\textwidth}{!}{
		\begin{tabular}{llrrrrrrr}
			Model                        & $\ell$ &      \texttt{Theft} &      \texttt{Other} &        \texttt{SBU} &      \texttt{Water} & \texttt{Wind} \& \texttt{Hail} &      \texttt{Fire} &       \texttt{Total} \\ \hline
			Baseline                     & 0      &               95.24 &              225.71 &              298.61 &              521.48 &                          46.74 &              47.28 &              1220.89 \\ \hline
			\multirow{3}{*}{ResNet18}    & 8      &           90.89 (0) & 234.88 (\textbf{1}) & 295.44 (\textbf{4}) &          521.16 (0) &             46.86 (\textbf{1}) & 53.01 (\textbf{2}) & 1227.32 (\textbf{2}) \\
			                             & 16     &           95.16 (0) & 226.80 (\textbf{1}) & 295.27 (\textbf{5}) & 522.05 (\textbf{2}) &             51.90 (\textbf{2}) & 55.51 (\textbf{2}) & 1227.56 (\textbf{3}) \\
			                             & 32     &          103.94 (1) & 229.50 (\textbf{2}) & 292.98 (\textbf{5}) & 529.70 (\textbf{3}) &             59.47 (\textbf{3}) &                 NA & 1236.52 (\textbf{6}) \\ \hline
			\multirow{3}{*}{ResNet50}    & 8      &           90.07 (0) & 234.23 (\textbf{1}) & 299.18 (\textbf{4}) &          523.22 (0) &             46.38 (\textbf{1}) & 44.54 (\textbf{2}) & 1224.86 (\textbf{2}) \\
			                             & 16     &  89.60 (\textbf{1}) & 235.49 (\textbf{1}) & 297.73 (\textbf{5}) & 521.87 (\textbf{1}) &             52.65 (\textbf{2}) & 81.20 (\textbf{5}) & 1224.32 (\textbf{3}) \\
			                             & 32     &  96.21 (\textbf{4}) & 233.98 (\textbf{2}) & 298.03 (\textbf{5}) & 530.90 (\textbf{2}) &             67.74 (\textbf{5}) & 75.43 (\textbf{4}) & 1224.57 (\textbf{6}) \\ \hline
			\multirow{3}{*}{ResNet101}   & 8      &  92.97 (\textbf{1}) &          226.25 (0) & 300.94 (\textbf{3}) & 525.15 (\textbf{1}) &             46.60 (\textbf{1}) & 57.26 (\textbf{1}) & 1230.82 (\textbf{2}) \\
			                             & 16     &  94.56 (\textbf{1}) &          227.19 (0) & 300.49 (\textbf{3}) &          522.54 (0) &             56.20 (\textbf{2}) &                 NA & 1229.13 (\textbf{2}) \\
			                             & 32     &           97.35 (1) & 224.73 (\textbf{3}) & 300.31 (\textbf{5}) &          521.75 (1) &             69.21 (\textbf{5}) & 76.24 (\textbf{2}) & 1227.08 (\textbf{4}) \\ \hline
			\multirow{3}{*}{DenseNet121} & 8      &  94.61 (\textbf{3}) & 231.82 (\textbf{1}) & 298.49 (\textbf{2}) &          520.42 (0) &             47.21 (\textbf{1}) & 50.15 (\textbf{1}) & 1229.60 (\textbf{2}) \\
			                             & 16     &  98.96 (\textbf{2}) & 231.40 (\textbf{2}) & 298.98 (\textbf{2}) & 518.73 (\textbf{1}) &             51.02 (\textbf{2}) &                 NA & 1233.66 (\textbf{1}) \\
			                             & 32     & 101.26 (\textbf{4}) & 224.64 (\textbf{4}) & 306.69 (\textbf{2}) &          518.62 (1) &             58.70 (\textbf{5}) & 68.30 (\textbf{6}) & 1233.37 (\textbf{2}) \\ \hline
		\end{tabular}}
	\caption{Testing deviance for severity prediction with principal components.}\label{tab:results-sev-pca}
\end{table}

Overall, the results for the severity models are much less impressive than those for the frequency models. The only exception is for the \texttt{SBU} peril, where one mostly observes a slight reduction in deviance and where unsupervised transfer learning improves the results compared with PCA. In general, one could conclude that the embeddings do not contribute much to the baseline severity model. One reason may be the limited dataset size (some perils have under 100 observations) or the embeddings do not capture useful features for severity modelling. 

\section{Discussion}\label{sec:conclusion}

We have proposed a simple model to use images within a ratemaking model. This approach does not drastically increase the number of parameters within the predictive model. We find that images improve the predictive ability of ratemaking models, meaning that there are observable characteristics within images that affect the risk of insurance contracts. 

We find statistically significant evidence for a relationship between image data and claim counts for certain perils. This is not the case for claim severity models. Our approach relies on embeddings, so we cannot conclude that there is a causal relationship between the image data and the claims count data. Our work, however, hints that there is probably a causal effect that can be identified in images. Future work could investigate which parts of the image have causal impacts on premiums, such that insurance companies could start collecting this information to use within pricing models. 

We also note that we considered other sources of images to attempt to answer our research question. First, we tried to use images from real estate websites. The advantages of such images are that they are of high quality, the pictures of facades are well framed in the image, and we have access to much-structured information that one could use for fine-tuning the image models to related tasks. A disadvantage of this approach is that a limited number of houses have a real-estate listing. If an insurance company attempts to use images to provide a quote to a potential customer, the image of that customer's home could not be available from one of the real-estate websites. Therefore, we needed to use a data source available for most of the potential customers in a region. Then, we considered using aerial imagery (for instance, Google Satellite). Note that most of these images for residential areas are not taken by satellites but by planes flying at low altitudes with high-resolution cameras. The advantage of this approach is that aerial imagery is available for most cities on Google Satellite. Otherwise, an insurance company can access a data provider for higher-quality data imagery. See, for instance, \cite{liu2023review} for a review of CNNs to aerial imagery, including applications such as object detection, classification and semantic segmentation. We decided not to use aerial imagery since we have not found useful, structured information related to insurance such that we may fine-tune the image representations to insurance-related tasks. However, we believe that aerial imagery will become an essential tool for risk assessment and insurance pricing since one could extract useful information such as land area, house area, presence or not of additions such as garages or pools, or the presence or not of objects that could cause claims such as trees or electric poles. 

Our goal in this research was to provide empirical evidence of useful information in SVI for insurance pricing. Therefore, this exploratory work serves as a proof-of-concept for this ratemaking framework. For this reason, we construct the embeddings using data freely available online for anybody to access. With the demonstrated usefulness of image data for insurance ratemaking, insurance companies might consider investing additional time and money in collecting and using this data source. For instance, one could attempt to obtain better quality images by building their dataset of SVI by taking individual pictures of houses that are all high-quality and centred on the house of interest. Also, we fine-tuned our image models on tax assessment datasets, but insurance companies already have access to features of interest for homes they insure since they collect data about the house during the quoting process. Therefore, one could construct representations of images using more insurance-related tasks. 

As a secondary objective of our project, we showed that one could use SVI to extract useful information about houses using online data automatically. Since our fine-tuning data was limited, we could only extract information like the construction year, the number of stories and the land/building/total value. However, insurance companies have access to internal data collected during the quoting process, such as the roofing material, the facade material and the presence or not of garages, sheds or pools. Therefore, another use of our framework is automatically extracting features from online images. In that case, insurance companies could attempt to predict some characteristics of the insured house without asking the customer to fill out the information manually. For instance, if the image model is confident that the facade type is made of a specific material, that field in the quoting form could be pre-filled with that material, improving the customer experience. 

The representation learning framework is beneficial since it lets us empirically verify that SVI contains valuable information to predict the claim frequency of home insurance contracts. However, since image models include so many parameters (even the smallest model we consider has over 6 million parameters in the convolutional weights alone), they are prone to overfit on the predictive task. For this reason, we train the model on related tasks such that the associated representations are adapted to insurance-related tasks and avoid overfitting on frequency or severity prediction. 

The objective of these experiments was not necessarily to encourage insurance companies to use images within their ratemaking process but to show empirical evidence that useful information within images leads to better actuarial fairness. For example, we have shown that SVI is useful for predicting claim frequency for specific perils and works especially well for sewer backup claims. In future work, one should seek to identify the most informative parts of the image, providing insights into the underlying mechanisms that drive risk in home insurance. In that case, one would retain the increased actuarial fairness but could also interpret and communicate these results convincingly to management, regulators and customers, opening the door to more informed decision-making.

\section{Acknowledgments}

This work was partially supported by the Natural Sciences and Engineering Research Council of Canada (Blier-Wong: 559169, Marceau: 05605). We thank Intact Financial Corporation for the data, support, and comments from Eliane Belisle, Frédérique Paquet, and Étienne Girard-Groulx. We thank Cyril Blanc for his help with the dataset and Ronald Richman for an insightful comment. 

\bibliographystyle{apalike}
\bibliography{ref-img}

\end{document}